\documentclass[12pt]{article}
\usepackage{enumerate,amssymb}
\renewcommand{\a}{\alpha}
\renewcommand{\b}{\beta}
\newcommand{\g}{\gamma}           
\renewcommand{\d}{\delta}         
\newcommand{\e}{\varepsilon}

\newcommand{\ka}{\kappa}
\newcommand{\ki}{\chi}
\newcommand{\la}{\lambda}

\newcommand{\s}{\sigma}           
         
\newcommand{\f}{{\phi}}

\newcommand{\eps}{{\epsilon}}

\newcommand{\be}{\begin{equation}}
\newcommand{\ee}{\end{equation}}
\newcommand{\eqn}[1]{\label{#1}\end{equation}}

\newcommand{\bea}{\begin{eqnarray}}
\newcommand{\eea}{\end{eqnarray}}
\newcommand{\eqan}[1]{\label{#1}\end{eqnarray}}

\newcommand{\ba}{\begin{array}}
\newcommand{\ea}{\end{array}}

\newcommand{\nn}{\nonumber}

\begin{document}

\begin{center}
{\bf   D=10, N=1 Supergravity and Second Order String Corrections}\\[14mm]

D. O'Reilly \\

{\it Physics Department\\ The Graduate School and University Center\\365 Fifth Avenue\\
New York, NY 10016-4309}\\[6mm]

\end{center}
\vbox{\vspace{3mm}}

\begin{abstract}

We study the Low Energy Limit of string theory in the form of string
corrected Supergravity. We do this in the non minimal case. That is
we solve the case of string corrected, D=10, N=1 Supergravity at
second order in the string slope parameter. We obtain the second
order H sector tensors, as well as the related torsions and
curvatures. We also find the supercurrent supertensor at second
order.\vbox{\vspace{1mm}}

\end{abstract}

\begin{center}
{  This work forms part I of a thesis submitted in partial
fulfilment of the requirements for the degree of Doctor Of Philosophy}\\[14mm]

\end{center}

\newpage

\begin{center}

{\Large Non Minimal String Corrections and Supergravity}

\end{center}

\hrule \tableofcontents \vspace{0.5cm}

\hrule

\vspace{0.5cm}

\newpage


\section{Introduction}

It is established tradition that theories of extended objects have
eclipsed supergravity as a focus of attention in the search for a
consistent theory of all known interactions. However in the past,
common ground for both theories was found, (see references \cite{1}
through \cite{5}, and references therein.) It is believed that D=10,
N=1 Supergravity is in fact the low energy limit to string theory,
\cite{1}. The first order or minimal Lorentz Chern-Simmons Form
string modifications or deformations to D=10, N=1 Supergravity, that
was manifestly supersymmetric, was first completed in \cite{4}. That
is, the authors succeeded in closing the related Bianchi identities
to first order in the string slope parameter, obtaining the minimal
string corrections. The completion of the task at second order or in
the non-minimal case has remained unsolved until now. It was not
known how to close the Bianchi identities at second order. Also the
relevance of such a lengthy task was not fully realized until
recently when the first order case was revisited. The results of
\cite{4} have recently been strongly vindicated, (\cite{2} and
references therein). Hence there has been a renewed focus of
interest in the non-minimal case, [2].

In view of this new interest in string corrected D=10, N=1
Supergravity, we therefore revisit an outstanding problem concerning
the case at second order in the perturbative expansion.
 Some years ago a program to incorporate string corrections into the supergravity equations of
motion which succeeded in maintaining manifest supersymmetry was
developed, \cite{1}, \cite{2}, \cite{3}, \cite{4}. Recently the
bosonic equations of motion for D=10, N=1 supergravity fields at
superspace and component levels have been derived and have been
shown to be derivable from a lagrangian, \cite{2}. This was done to
first order perturbatively in the string slope parameter. A major
problem continued to be that of obtaining the solution of the
Bianchi identities at second order. It was suggested, \cite{4}, that
the second order case would contain many interesting results, and
that it would follow by merely obtaining perturbative expansions of
the super tensors, $H_{abc}$, $L_{abc}$, $A_{abc}$. However
obtaining the second order result did not come so easily. (Some
authors maintained a different approach to that in \cite{2}, see for
example \cite{5}).

Closing the Bianchi identities at second order first ground to a
halt upon encountering a non-solvable term in the H sector Bianchi
identities. Many approaches such as null spaces and other ideas were
tried in order to overcoming this problem.

It was suggested that one aspect of finding a second order solution
would require the so called X tensor, \cite{1}, \cite{7}. Finding
the X tensor would form part of the solution, and finding it through
direct calculations proved difficult. However this author was aware
for a long time that a particular Ansatz did in fact allow for
closure in the H sector,(equation (69)). However it was required to
find a consistent set of torsions and curvatures also. Eventually,
an equation was derived which allowed for a solution up to a
curvature, equation (39). The remaining curvature looked intractable
until yet a further relation was found, (equation (124)). Finally a
condition to be imposed on the curvature $R_{ab}{}_{\a}{}^{\b}$
(equation 129), which was not evident in referenc \cite{4}, was
implemented, (129).

Hence in this work we propose a candidate for the X tensor and show
that it allows for a solution in the H sector. Furthermore we solve
the related torsion identities. We also find $R^{(2)}{}_{\a\b de}$,
$R^{(2)}{}_{\a bde}$. We find  $R^{(1)}{}_{ab}{}_{\g}{}^{\d}$ and
note that it appears to be set to zero in \cite{2} and \cite{4}.
Finally we find $A^{(2)}{}_{abc}$, the super current super tensor at
second order.

The perturbative approach of Gates and collaborators is well
documented and discussed in the literature, and we will not recount
it here. For a recent review and for an up to date commentary see
\cite{2}, and references therein. For a discussion of Bianchi
identities in general see \cite{6}. The crucial role of the
Chern-Simmons form is discussed in \cite{8} and \cite{4}. Our
starting point will be the Bianchi identities as listed in \cite{4}.
The sigma matrix identities and symmetries are recorded in \cite{3}.

These geometrical methods nowadays are known as deformations,
\cite{2}, and the constraints have sometimes been referred to in the
past as Beta Function Favored constraints, ($\b FF $ constraints).

\section{The Bianchi Identities}
 Bianchi Identites in the presence of constraints can give
 information about dynamics. Perhaps the best known example in
 physics is the case of the Einstein Field Equations. It is well known that
 these field equations can be derived from an action, the Einstein
 Hilbert action. However, as is also well known, the standard route
 towards their derivation is through the contraction of the Bianchi
 identities in classical Riemannian space. An example of a physical
 constraint is that of the conservation of the energy momentum
 tensor. Hence although the Bianchi identities are in fact just
 identities, if a constraint is imposed on one or more fields, this
 in turn will generate constraints on other fields related through
 these identities.

In our case we consider in a slightly analogous way, those Bianchi
identities derivable in superspace in conjunction with a set of
constraints. For details of the early work on this see for example
\cite{1} and references therein.  Following the notation of \cite{6}
we define torsions $T_{AB}{}^{C}$ and curvatures $R_{AB}{}^{CD}$ as
follows

\bea
[\nabla_{A},\nabla_{B}\}=T_{AB}{}^{C}+\frac{1}{2}R_{ABd}{}^{e}M_{e}{}^{d}
\eea

The Bianchi identities are given by

\bea[[\nabla_{[A},\nabla_{B}\},\nabla_{C)}\}=0\eea

For further details see appendix [II].
 Firstly we consider the set of Q and G Bianchi
identities is listed in ref. \cite{4}. $Q_{ABD}$ is the Super
Lorentz Chern Simmons Form. $G_{ABD}$ is the supergravity field
strength for D=10, N=1 Supergravity.  $H_{ABD}$ will be the modified
supergravity field strength for D=10, N=1 Supergravity

The Bianchi identities satisfied by the Lorentz Chern Simmons form,
$Q_{ABD}$, are \cite{4}

\bea
\frac{1}{6}\nabla_{(\a|}Q_{|\b\g\d)}-~\frac{1}{4}T_{(\a\b|}{}^{E}
Q_{E|\g\d)}=~\frac{1}{4}R_{(\a\b|ef}R_{|\g\d)}{}^{ef} \eea

 \bea
{\frac{1}{2}}\nabla_{(\a|}Q_{|\b\g)d}-{\nabla_{d}}Q_{\a\b\g}~
-~\frac{1}{2}T_{(\a\b|}{}^{E}Q_{E|\g)d}~
+~\frac{1}{2}T_{d(\a|}{}^{E}Q_{E|\b\g)}~=~R_{(\a\b|ef}R_{|\g)d}{}^{ef}
\eea

 \bea~ \nabla_{(\a|}Q_{\b)
cd}~+~\nabla_{[c|}Q_{|d]\a\b}~-~T_{\a\b}{}^{E}Q_{Ecd}~
-~T_{cd}{}^{E}Q_{E\a\b}~-~T_{(\a|[c|}{}^{E}Q_{E|d]|\b)}~=~\nn\\
~[2R_{\a\b ef}R_{cd}{}^{ef}~
+~R_{(\a|[c|}{}^{ef}R_{|d]|\b)}{}_{ef}]\eea

\bea
\nabla_{\a}Q_{bcd}-\frac{1}{2}\nabla_{[b|}Q_{\a\|cd]}-\frac{1}{2}T_{\a[b|}{}^{E}Q_{E|d]\a}
+\frac{1}{2}T_{[bc|}{}^{E}Q_{E|d]\a}=R_{\a[b|}{}^{ef}R_{|cd]ef}\nn\\
\eea\ \bea
+\frac{1}{6}\nabla_{[a|}Q_{|bcd]}-\frac{1}{4}T_{[ab|}{}^{E}Q_{E|d]\a}
+\frac{1}{2}T_{[bc|}{}^{E}Q_{E|cd]}=\frac{1}{4}
R_{[ab|}{}^{ef}R_{|cd]ef} \eea

Those satisfied by the supertensor, $G_{ABC}$, are, \cite{4},
 \bea
\frac{1}{6}\nabla_{(\a|}G_{|\b\g\d)}-~\frac{1}{4}T_{(\a\b|}{}^{E}
G_{E|\g\d)}=0 \eea
 \bea
{\frac{1}{2}}\nabla_{(\a|}G_{|\b\g)d}-{\nabla_{d}}G_{\a\b\g}~
-~\frac{1}{2}T_{(\a\b|}{}^{E}G_{E|\g)d}~
+~\frac{1}{2}T_{d(\a|}{}^{E}G_{E|\b\g)}~=~0 \eea

 \bea~ \nabla_{(\a|}G_{\b)
cd}~+~\nabla_{[c|}G_{|d]\a\b}~-~T_{\a\b}{}^{E}G_{Ecd}~
-~T_{cd}{}^{E}G_{E\a\b}~-~T_{(\a|[c|}{}^{E}G_{E|d]|\b)}~=0~\nn\\
\eea

\bea
\nabla_{\a}G_{bcd}-\frac{1}{2}\nabla_{[b|}G_{\a\|cd]}-\frac{1}{2}T_{\a[b|}{}^{E}G_{E|d]\a}
+\frac{1}{2}T_{[bc|}{}^{E}G_{E|d]\a}=0\nn\\
\eea\

\bea
+\frac{1}{6}\nabla_{[a|}G_{|bcd]}-\frac{1}{4}T_{[ab|}{}^{E}G_{E|d]\a}
+\frac{1}{2}T_{[bc|}{}^{E}G_{E|cd]}= 0\eea

The $Q$ and $G$ tensors are related the modified field strength
tensor, the $H$ tensor, in superspace as follows \bea G_{ADG}~=
H_{ADG}~+~\g Q_{ADG}+ \b Y_{ADG} \eea
 $Y_{ADG}$ is the Yang
Mills Superform, and $\g$ is proportional to the string slope
parameter. That is, the action for massless fields of heterotic or
type I superstrings may be expanded, with $\b$ set to zero as
follows, [1],
 \bea S_{eff}~=~\frac{1}{\ka^{2}}\int
d^{10}x
e^{(-1)}[\emph{L}_{(0)}~+~\sum_{n=1}^{n=\infty}(\gamma')^{n}\emph{L}_{(n)}]
\eea

Thus we can arrive at the low energy string corrected effective
action.

 We now consider the Bianchi identities satisfied by the tensor $H_{ABC}$ which
 are as followes

 \bea
\frac{1}{6}\nabla_{(\a|}H_{|\b\g\d)}-~\frac{1}{4}T_{(\a\b|}{}^{E}
H_{E|\g\d)}=~(-\frac{\g}{4})R_{(\a\b|ef}R_{|\g\d)}{}^{ef} \eea
 \bea
{\frac{1}{2}}\nabla_{(\a|}H_{|\b\g)d}-{\nabla_{d}}H_{\a\b\g}~
-~\frac{1}{2}T_{(\a\b|}{}^{E}H_{E|\g)d}~
+~\frac{1}{2}T_{d(\a|}{}^{E}H_{E|\b\g)}~=~(-\g)R_{(\a\b|ef}R_{|\g)d}{}^{ef}
\eea

 \bea~ \nabla_{(\a|}H_{\b)
cd}~+~\nabla_{[c|}H_{|d]\a\b}~-~T_{\a\b}{}^{E}H_{Ecd}~
-~T_{cd}{}^{E}H_{E\a\b}~-~T_{(\a|[c|}{}^{E}H_{E|d]|\b)}~=~\nn\\
~-\g[2R_{\a\b ef}R_{cd}{}^{ef}~
+~R_{(\a|[c|}{}^{ef}R_{|d]|\b)}{}_{ef}]\eea

\bea
\nabla_{\a}H_{bcd}-\frac{1}{2}\nabla_{[b|}H_{\a\|cd]}-\frac{1}{2}T_{\a[b|}{}^{E}H_{E|d]\a}
+\frac{1}{2}T_{[bc|}{}^{E}H_{E|d]\a}=-\g R_{\a[b|}{}^{ef}R_{|cd]ef}\nn\\
\eea\

\bea
+\frac{1}{6}\nabla_{[a|}H_{|bcd]}-\frac{1}{4}T_{[ab|}{}^{E}H_{E|d]\a}
+\frac{1}{2}T_{[bc|}{}^{E}H_{E|cd]}=-\frac{1}{4}\g
R_{[ab|}{}^{ef}R_{|cd]ef} \eea

Within the framework of the Bianchi identities we have a
perturbative prescription that will allow us to incorporate string
corrections into the theory, and maintain it manifestly
supersymmetric. We first solve the identities satisfied by the H
tensor. To find them we use the Bianchi identities as satisfied by
the $G$ tensor and the Lorentz Chern Simmons form as listed in
\cite{4}. We then combine them to get the H sector Bianchi
Identities. We also have the torsions

\bea
T_{(\a\b|}{}^{\la}T_{|\gamma)\la}{}^{d}~-~T_{(\a\b|}{}^{g}T_{|\gamma)g}{}^{d}~
-~\nabla_{(\a|}T_{\b\g)}{}^{d}~=~0 \eea
 \bea
T_{(\a\b|}{}^{\la}T_{|\g)\la}{}^{\d}~-~T_{(\a\b|}{}^{g}T_{|\g)g}{}^{\d}
~-~\nabla_{(\a|}T_{|\b\g)}{}^{\d} -~\frac{1}{4}R_{(\a\b|
de}\s^{de}{}_{|\g)}{}^{\d} ~=~0 \eea

And we also have the following curvature

\bea
T_{(\a\b|}{}^{\la}R_{|\g)\la}{}_{de}~-~T_{(\a\b|}{}^{g}R_{|\g)}{}_{gde}~
-~\nabla_{(\a|}R_{\b\g)}{}_{de}~=~0\eea

There are many other identities but these are the principle ones
needed for this work. In fact we only need to solve the first three
$H$ sector identities to find a full solution for the H sector
tensors in terms of curvatures and torsions.

 The first order solutions were first given in [4]. They were
recently recalculated  [2]. Using conventional constraints as input,
Bellucci, Gates and Depireux derived the first order results, as
given in [4]. The conventional constraints, (where here we use a
slightly different convention) are, \cite{2},

\bea i\s_{a}{}^{\a\b}T_{\a\b}{}^{b}=
16\d_{a}^{b}~~~,i\s_{c}{}^{\a\b}T_{\a b}{}^{c}=0\nn\\
i\s_{abcde}{}^{\a\b}T_{\a\b}{}^{e}=0,~~~~~~~T_{\a[de]}=0\nn\\
T_{deb}=\frac{1}{8}\s_{de}{}_{\a}{}^{\b}T_{\b
b}{}^{\a}-\frac{i}{16}R_{\a\b~de}\nn\\
\eea

An important input at first order is what is taken to be the
supercurrent supertensor $A_{abc}$.

The choice with $\b$ set to zero of

\bea A_{abc}=-i\g\s_{abc \eps\tau}
T_{kp}{}^{\eps}T^{kp}{}^{\tau}\eea was made to put the theory on
shell,  \cite{2}.

In this work we also find $A_{abc}$ at second order. We will list
the main results at first order found in \cite{2} and \cite{4}. This
is because we continually refer to them. We also wish to establish
out notation, and because we need these results in terms of the $H$
tensor, not the $G$ tensor as given in \cite{4}. We list only those
which we require, and we set $\b =0$. (ie we include no matter
fields). We will note however that we will later have to modify the
constraint on $T_{\a\b}{}^{g}$.

We have at first order \bea H_{\a\b\g}=0~~~~~H_{\a\b
d}=+\frac{i}{2}\s_{d \a\b}
+4i\g\s^{g}{}_{\a\b}H_{\g}{}^{ef}H_{d}{}^{ef}\nn\\
H_{\a bd}=+2i\g[-\s_{[b|}{}_{\a\b}T_{ef}{}^{\b}G_{|d]}{}^{ef}-
2\s_{e}{}_{\a\b}T_{f[b|}{}^{\b}G_{|d]}{}^{ef}]\nn\\
\eea \bea T_{\a\b}{}^{g}= i\s_{\a\b}{}^{g}, ~~~T_{\a b}{}^{g}=0,
~~T_{abc}=-2L_{abc}, \nn\\
T_{\a\b}{}^{\g}=-[\d_{(\a|}{}^{\g}\d_{|\b)}{}^{\d}+\s^{g}{}_{\a\b}\s_{g}{}^{\g
\d}]\chi_{\d}\nn\\
T_{\a b}{}^{\g} =
 \frac{1}{48}\s_{b \a \f}\s^{pqr}{}^{\f\g}A_{pqr}\nn\\
\eea \bea R_{\a\b de} =
-2i\s^{g}{}_{\a\b}\Pi_{gde}+\frac{i}{24}\s^{pqref}{}_{\a\b}A_{pqr}\nn\\
 R_{\a bde}= -i\s_{[d|}{}_{\a \f}T_{b |e]}{}^{f} +i\g\s_{[d|}{}_{\a~
 \f}T_{kl}{}^{\f}R^{kl}{}_{|de]}\nn\\
 \eea
 Where

 \bea
 \Pi^{(1)}{}_{g}{}^{ef}=L^{(1)}{}_{g}{}^{ef}
-~\frac{1}{8}A^{(1)}{}_{g}{}^{ef} \eea

\bea L_{abc}= H_{abc}+\g[(R^{ef}{}_{|ab]}+
R_{|ab]}{}^{ef}-\frac{8}{3}H_{d}{}^{ef}{}_{[a|}H^{df}{}_{|b]})H_{|c]ef}]\eea

\bea \nabla_{\a}L_{bcd}=\frac{i}{4}\s_{[b|\a\b}[T_{|cd]}{}^{\b}+4\g
T_{kl}{}^{\b}R^{kl}{}_{|cd]}\nn\\
\nabla_{\g}T^{(0)}{}_{ef}{}^{\d}=[-\frac{1}{4}\s^{mn}{}_{\g}{}^{\d}R_{efmn}
~+~T^{(0)}{}_{ef}{}^{\la}T^{(0)}{}_{\g\la}{}^{\d}] \nn\\
+\frac{1}{48}[2H_{efg}\s^{g}{}_{\g}{}^{\f}\s^{dkl}{}_{\f}{}^{\d}-\s_{[e|}{}_{\g}{}^{\f}\s^{dkl}{}_{\f}{}^{\d}\nabla_{|f]}]A_{dkl}
\eea

Many avenues such as null spaces were tried in order to solve the
problem of finding a solution at second order, but without success.
However it was suggested that a generalization of the torsion
$T_{\a\b}{}^{g}$ would be necessary in order to proceed to second
order, \cite{2},\cite{7}. The job at hand therefore is to find the
form of this generalization, known as the X tensor. However there
were still many obstacles to be overcome in order to obtain a
complete and consistent set of solutions.

In this paper we propose a candidate for the X tensor. We show that
this candidate solves the problem of closure in the H sector Bianchi
Identities. It also solves the second order torsion. We therefore
find  the second order torsions $T^{(2)}{}_{\a\b}{}^{g}$,
$T^{(2)}{}_{\a\b}{}^{\la}$, $T^{(2)}{}_{\a b}{}^{g}$ and
$T^{(2)}{}_{\a b}{}^{\g}$ and also the curvatures $R^{(2)}{}_{\a
bde}$, and $R^{(2)}{}_{\a\b de}$.  In order to do this we required
several insights and key results which are derived in appendices.

We check our results by showing mutual consistency. At this stage we
will draw attention to our simple second order notation. The
superscript in brackets refers to the second order quantities. Hence
we have as follows for example

\bea R_{\a\b de} = R^{(0)}{}_{\a\b de} + R^{(1)}{}_{\a\b de}+
R^{(2)}{}_{\a\b de} +....\eea

Where $R^{(0)}{}_{\a\b de}$ and $R^{(1)}{}_{\a\b de}$ are listed in
equations (27). To begin with, $H_{\b\g\d}$ is set to zero as in
\cite{2} and \cite{4}. We have not seen that it is required to be
other than zero to close the Bianchi identities. We have seen that
if it is non zero the H sector Bianchi Identities fail to close.

\section{The X Tensor}
In reference \cite{4} the conventional constraint
$T^{g}{}_{\a\b}~=~i\s^{g}{}_{\a\b}$ was imposed to all orders. This
led to failure to close the Bianchi identities at second order. In
this case that constraint is modified. From the conventional
constraints listed in \cite{2} the most general form of the zero
dimensional torsion is
 \bea T_{\a\b}{}^{g}~=~i\s^{g}{}_{\a\b}
~+~\s^{pqref}{}_{\a\b}X_{pqref}{}^{g} \eea
 Here we absorb the coefficient $\frac{i}{5!}$ used in ref. [2] into the X tensor.

 Earlier because of the existence of an apparently intractable
term which arose in the H sector Bianchi identities, closure could
not be obtained with $T_{\a\b}{}^{g}=i\s^{g}{}_{a\b}$. At the time
the problem term could not be incorporated into the torsion
$T^{(2)}{}_{\a\b}{}^{\la}$, which would have allowed for a solution.
In the following we see that the X tensor must contribute to second
order.
 Consider the Bianchi identity at dimension one half, equation (20).
 If X is zero, then
using the constraints in \cite{4}, reduces this equation to \bea
T^{(2)}{}_{(\a\b|}{}^{\la}\s_{|\gamma)\la}{}^{d}~=~\s_{(\a\b|}{}^{g}T^{(2)}{}_{|\gamma)g}{}^{d}
 \eea
 Therefore $T^{(2)}{}_{\a\b}{}^{\la}$ must have a similar sigma matrix structure to the RHS of (33).
 We know from the H sector Bianchi identities that
 $H^{(2)}{}_{g \g d}$ satisfies an equation of the form
\bea
\s^{g}{}_{(\a\b|}H^{(2)}{}_{g|\g)d}~~~=~~~~~~~~~~~~~~~~~~~~~~~~~~~~~~~~~~~~~~~~~~~~~~~~~\nn\\
\s^{g}_{(\a\b|}[M_{g|\g)d}]+~\s^{g}{}_{(\a\b|}T^{(2)}{}_{|\g) gd}
-~~\s_{d(\a|\la}T^{(2)}{}_{|\b\g)}{}^{\la}~
-~8\g^{2}{\s_{e(\a|\eps}}{\s_{f(\b|\tau}}{\s_{d|\g)\f}}
T^{(0)}{}_{kp}{}^{\eps}T^{(0)}{}^{kp}{}^{\tau}T^{(0)}{}^{ef}{}^{\f}\nn\\
~~~~~~~~~~~~~~~~~~~~~~~~~~~~~\eea Looking at equation (34) we see
that we will encounter an intractable term in the H sector unless we
can absorb it into the $T^{(2)}{}_{\b\g}{}^{\la}$ term, that is the
fourth term on the RHS of (34). Let us call it the $T^{3}$ term.
There is no known sigma matrix identity that will allow this term to
be written in the form necessary to solve for $H_{g \g d}$. We see
that the $T^{3}$ term has the same sigma matrix structure as the
$T^{(2)}{}_{\b\g}{}^{\la}$ term. We consider the option of equating
the $T^{(2)}{}_{\b\g}{}^{\la}$ term with the $T^{3}$ term. This is
possible only when X is non zero, and it will constitute one of the
many necessary components of the solution.

\section{Results Necessary For Solution}

In order to see relationships and recognize cancelations we retain
on our simplified index notation, for example
$T^{(2)}{}_{\b\g}{}^{\la}$, and keep in mind that quantities can be
written in several ways which make them recognizable. We employ the
following results which are crucial to obtaining a solution. We will
require the zeroth order spinor derivative of
$T^{(0)}{}_{ef}{}^{\d}$, available from (30),

\bea
\nabla_{\g}T^{(0)}{}_{ef}{}^{\d}=[-\frac{1}{4}\s^{mn}{}_{\g}{}^{\d}R_{efmn}
~+~T^{(0)}{}_{ef}{}^{\la}T^{(0)}{}_{\g\la}{}^{\d}] \nn\\\eea

 We refer to its $\chi$ part and its curvature part using
 a suitable subscript as in (36).

We have the following results, the derivations of which are lengthy
and are included in appendices.

\bea
~T^{(0)}{}_{(\a\b|}{}^{\la}\frac{i\g}{6}\s^{pqref|}{}_{\g)\la}A^{(1)}{}_{pqr}H^{(0)}{}_{def}
-\frac{i\g}{6}\s^{pqref}{}_{(\a\b|}[H^{(0)}{}_{def}\nabla_{|\g)}A^{(1)}{}_{pqr}]\mid_{(\chi)}\nn\\
=~+~i\g\s^{g}_{(\a\b|}H^{(0)}{}_{def}\nabla_{|\g)}A^{(1)}{}_{gef}\mid_{(\chi)}
\eea We also require, (See appendix IV.)

 \bea
+~\frac{i\g}{6}\s^{pqref}{}_{(\a\b|}H^{(0)}{}_{def}\nabla_{|\g)}A^{(1)}{}_{pqr}|_{(R)}
=~ \frac{\g^{2}}{12}\s^{pqref}{}_{(\a\b|}H^{(0)}{}_{def}
\s_{pqr}{}_{\eps\tau}T^{(0)}{}_{kp}{}^{\eps}\s^{mn}{}_{|\g)}{}^{\tau}R^{(0)}{}^{kp}{}_{mn}\nn\\
=~\frac{\g^{2}}{2}\s^{g}{}_{(\a\b|}\s_{gef}{}_{\eps\tau}T^{(0)}{}_{kp}{}^{\eps}\s^{mn}{}_{|\g)}{}^{\tau}
R^{(0)}{}^{kp}{}_{mn}H^{(0)}{}_{d}{}^{ef}
+~16\g^{2}\s^{g}{}_{(\a\b|}\s_{e|\g)\eps}T^{(0)}{}_{kp}{}^{\eps}R^{(0)}{}_{fg}{}^{kp}H^{(0)}{}_{d}{}^{ef}\nn\\
~\eea
 And for later transparency we have the very important and
 convenient observation that,  using the identity
 $\s^{g}{}_{(\a\b|}\s_{g |\g)\la}=0$, this can be written as
\bea
~\frac{i\g}{6}\s^{pqref}{}_{(\a\b|}H^{(0)}{}_{def}\nabla_{|\g)}A^{(1)}{}_{pqr}|_{(R)}\nn\\
=-i\g\s^{g}{}_{(\a\b|}[\nabla_{|\g)}A^{(1)}{}_{gef}]|_{(R)}H^{(0)}{}_{d}{}^{ef}+
4i\g \s^{g}{}_{(\a\b|}R^{(1)}{}_{|\g) gef}H^{(0)}{}_{d}{}^{ef}\eea

Combining the results (36) and (38) we get the very simple
reduction, \bea
~+\frac{i\g}{6}T^{(0)}{}_{(\a\b|}{}^{\la}\s^{pqref}{}_{|\g)\la}A^{(1)}{}_{pqr}H^{(0)}{}_{def}-
\frac{i\g}{6}\s^{pqref}{}_{(\a\b|}H^{(0)}{}_{def}\nabla_{|\g)}A^{(1)}{}_{pqr}\nn\\
=+i\g\s^{g}{}_{(\a\b|}[\nabla_{|\g)}A^{(1)}{}_{gef}]H^{(0)}{}_{d}{}^{ef}-
4i\g\s^{g}{}_{(\a\b|}R^{(1)}{}_{|\g) gef}H^{(0)}{}_{d}{}^{ef}\eea

This will be a key part of the solution. Along with the sigma matrix
identities given in \cite{3}, we also require the important result
that
 \bea
\s^{pqref}{}_{(\a\b|}\s_{e|\g)\f}=~=~-~\s^{pqref}{}_{\f(\g|}\s_{e|\a\b)}
\eea The latter allows us to write

\bea \frac{i\g}{12}\s^{pqref}{}_{(\a\b|}A^{(1)}{}_{pqr}
R^{(0)}{}_{|\g)}{}_{def}=~+\frac{\g}{6}\s^{g}{}_{(\a\b|}\s^{pqre}{}_{g}{}_{|\g)\f}A^{(1)}{}_{pqr}T^{(0)}{}_{d
e}^{}{\f} \eea

Hence we transform it to a solvable term. It is the above system of
derived identities and observations, coupled with an Anzatz for the
X tensor that will facilitate our solution up to a curvature.
Solving the final curvature will involve other technical
difficulties.

\section{The H Sector Solution}
 To second order in perturbation theory we have, using our
 exponent notation, for  $H^{(2)}{}_{\b\g d}$
\bea
\frac{1}{6}\nabla_{(\a|}H_{|\b\g\d)}{}^{Order(2)}~-~\frac{1}{4}T^{(0)}{}_{(\a\b|}{}^gH^{(2)}{}_{g|\g\d)}~
-~~\frac{1}{4}T^{(1)}{}_{(\a\b|}{}^gH^{(1)}{}_{g|\g\d)}~-~\frac{1}{4}T^{(2)}{}_{(\a\b|}{}^gH^{(0)}{}_{g|\g\d)}~\nn\\
-~~\frac{1}{4}T^{(0)}{}_{(\a\b|}{}^{\la}H^{(2)}{}_{\la|\g\d)}~-~\frac{1}{4}T^{(1)}{}_{(\a\b|}{}^{\la}H^{(1)}{}_{\la|\g\d)}~
-~~\frac{1}{4}T^{(2)}{}_{(\a\b|}{}^{\la}H^{(0)}{}_{\la |\g\d)}~ =
~-\frac{\g}{2}R^{(1)}{}_{(\a\b|ef}R^{(0)}{}_{|\g\d)}{}^{ef}
\nn\\\eea
 For
$H^{(2)}{}_{ \b g d}$ we have

 \bea {\frac{1}{2}}\nabla_{(\a|}H_{|\b\g)d
}^{Order\g^{2}}~~- ~~\nabla_{d}H_{\a\b\g}^{Order\g^{2}}
~~-~~\frac{1}{2}T^{(0)}{}_{(\a\b|}{}^{\la}H^{(2)}{}_{\la|\g)d}~~
-~~\frac{1}{2}T^{(1)}{}_{(\a\b|}{}^{\la}H^{(1)}{}_{\la|\g)d}~~\nn\\
-~~\frac{1}{2}T^{(2)}{}_{(\a\b|}{}^{\la}H^{(0)}{}_{\la|\g)d}~
-~~\frac{1}{2}T^{(0)}{}_{(\a\b|}{}^{g}H^{(2)}{}_{g|\g)d}~~
-~~\frac{1}{2}T^{(1)}{}_{(\a\b|}{}^{g}H^{(1)}{}_{g|\g)d}~~
-~~\frac{1}{2}T^{(2)}{}_{(\a\b|}{}^{g}H^{(0)}{}_{g|\g)d}~~\nn\\
+~~\frac{1}{2}T^{(2)}{}_{d(\a}{}^{g}H^{(0)}{}_{g|\b\g)}~~
+~~\frac{1}{2}T^{(1)}{}_{d(\a}{}^{g}H^{(1)}{}_{g|\b\g)}
+~~\frac{1}{2}T^{(0)}{}_{d(\a}{}^{g}H^{(2)}{}_{g|\b\g)}\nn\\
+~~{\g'}[R^{(0)}{}_{(\a\b|ef}R^{(1)}{}_{|\g)d}{}^{ef}~~
+~~R^{(1)}{}_{(\a\b|ef}R^{(0)}{}_{|\g)d}{}^{ef}]~~=0 \nn\\\eea
 We need to
solve (42) and (43). We do so using the constraints of reference
\cite{4}.

It is a straightforward to solve for $H^{(2)}{}_{g\g\d}$. To do so
we use the constraints in ref. \cite{4} and substitute them into
equation (42). We extract a sigma matrix coefficient from each term
as below. After some care with the algebra we obtain.
 \bea
\s_{(\a\b|}{}^{g}H^{(2)}{}_{g|\g\d)}~~=~
~\s^{g}{}_{(\a\b|}[{-4\g}H^{(0)}{}_{gef}R^{(1)}{}_{|\g\d)}{}^{ef}~
~-~~\frac{1}{2}T^{(2)}{}_{g|\g\d)}] \eea

Which solves with the correct symmetries to simply

 \be H^{(2)}{}_{g\g\d}~~=~[{-4\g}H^{(0)}{}_{gef}R^{(1)}{}_{\g\d}{}^{ef}~
~-~~\frac{1}{2}T^{(2)}{}_{g \g\d}] \ee \
 Which we can also write as
 \bea
H^{(2)}{}_{d
\a\b}~=~\s_{\a\b}{}^{g}[{8i\g}H^{(0)}{}_{def}L^{(1)}{}_{g}{}^{ef}~
-~i\g H^{(0)}{}_{def}A^{(1)}{}_{g}{}^{ef}]\nn\\~+~
\s^{pqref}{}_{\a\b}[\frac{-i\g}{6}H^{(0)}{}_{def}A^{(1)}{}_{pqr}~-
~\frac{1}{2}X_{pqrefd}] \eea

For the next Bianchi Identity we will require considerably more
ingenuity. The result will also have to be extracted using a
symmetrization operator, equation (135). Applying the constraints of
ref. \cite{4} reduces (43) to

\bea {\frac{1}{2}}\nabla_{(\a|}H_{|\b\g)d}{}^{Order\g^{2}}~~
-~~\frac{1}{2}T^{(2)}{}_{(\a\b|}{}^{\la}H^{(0)}{}_{\la|\g)d}~~
-~~\frac{1}{2}T^{(0)}{}_{(\a\b|}{}^{\la}H^{(2)}{}_{\la|\g)d}~~
-~~\frac{1}{2}T^{(0)}{}_{(\a\b|}{}^{g}H^{(2)}{}_{g|\g)d}~~\nn\\
+~~\frac{1}{2}T^{(2)}{}_{d(\a}{}^{g}H^{(0)}{}_{g|\b\g)}~~
+~~{\g'}[R^{(0)}{}_{(\a\b|ef}R^{(1)}{}_{|\g)d}{}^{ef}~~
+~~R^{(1)}{}_{(\a\b|ef}R^{(0)}{}_{|\g)d}{}^{ef}]~~=0 \nn\\\eea

 We now substitute the constraints known from ref. \cite{4} into (47) as
before. After some long calculations and we obtain

\bea \frac{+i}{2}\s^{g}{}_{(\a\b|}H^{(2)}{}_{g|\g)d}~=~
{\frac{1}{2}}\nabla_{(\a|}H_{|\b\g)d}{}^{Order\g^{2}}
~-~\frac{i}{4}\s_{d(\a|\la}T^{(2)}{}_{|\a\b)}{}^{\la}\nn\\
~+~\s^{g}{}_{(\a\b|}[{-2\g}\s_{g}{}^{\la\f}\ki_{\f}H^{(0)}{}_{def}R^{(1)}{}_{\la|\g)}{}_{ef}~\nn\\
-~\frac{1}{4}\s_{g}{}^{\la\f}\ki_{\f}T^{(2)}{}_{d\la|\g)}]~
~-{4\g}\ki_{(\a|}H^{(0)}{}_{def}R^{(1)}{}_{|\b\g)}{~}^{ef}~\nn\\
-\frac{1}{2}\ki_{(\a}T^{(2)}{}_{d|\b\g)}
+~\frac{i}{4}\s^{g}_{(\a\b|}T^{(2)}{}_{d|\g)g}
{-2i\g}\s^{g}{}_{(\a\b|}H^{(0)}{}_{gef}R^{(1)}{}_{|\g)d}{~}^{ef}~~\nn\\
+~\s^{g}{}_{(\a\b|}[{-i2\g}
\Pi^{(1)}{}_{g}{}^{ef}R^{(0)}{}_{|\g)d}{}^{ef}]
+~\frac{i\g}{24}\s^{pqr}{}_{ef}{}_{(\a\b|}R^{(0)}{}_{|\g)d}{}^{ef}A^{(1)}{}_{pqr}
\eea Equation (48) contains a proliferation of non solvable non
linear terms. Hence we follow our first route. We consider the
spinor derivative of $H_{\b\g d}{}^{Order\g^{2}}$. Since we do not
yet know what the X tensor is we will first employ a torsion and a
curvature to eliminate $\chi$ and X tensor terms. We retain our
compact superscript notation for terms which will later cancel and
so we will not write them out in full unless required. We must also
remember to include the second order derivative contributions which
come from the first order result. In \cite{4} it was found to first
order,
 \bea H_{\b \g
d}~=~\frac{i}{2}\s_{d\b\g}+i4\g\s^{g}{}_{\b\g}H^{(0)}{}_{gef}H^{(0)}{}_{def}
\eea
 Taking the
spinor derivative of this first order term will generate second
order contributions. Hence we include this contribution. However we
do not list the terms explicitly. The contribution due to this term
is simply a lengthy expression with the correct sigma matrix
structure needed to extract the solution. The explicit result will
be written in full in a later paper. The method of solution for this
Bianchi identity involves extracting a sigma matrix similar to the
coefficient of the term which we seek. Namely we seek to solve for
$H^{(2)}{}_{g\g d}$ in an expression of the form

 \bea
\s^{g}{}_{(\a\b|}H^{(2)}{}_{g|\g)d}~=~\s^{g}{}_{(\a\b|}M^{(2)}{}_{g|\g)d}
\eea
 Finally we extract the correct expression for $H^{(2)}{}_{g\g d}$
with an appropriate symmetrization operator, (section (12)).
 We have from first order, which we leave as is,
\bea \frac{1}{2}\nabla_{(\a|}H^{(1)}{}_{|\b\g)
d}{}^{(Order\g^{2})}~~=~~
\s^{g}_{(\a\b|}({2i\g}\nabla_{|\g)}[H^{(0)}{}_{gef}H^{(0)}{}_{d}{}^{ef}])^{Order(1)}
\eea

We then have, taking the derivative of (45)

 \bea \frac{1}{2}\nabla_{(\a|}H^{(2)}{}_{|\b\g)
d}{}^{(Order\g^{2})}~=~
[{-2\g}\nabla_{(\a|}(H^{(0)}{}_{def}R^{(1)}{}_{|\b\g)}{}^{ef})~-~
\frac{1}{4}\nabla_{(\a|}T^{(2)}{}_{d|\b\g)}]~~~~~~~~ \eea
 \bea =
(\nabla_{(\a|}[{-2\g}H^{(0)}{}_{def}])R^{(1)}{}_{|\b\g)}{}^{ef}
~-~{2\g}H^{(0)}{}_{def}[\nabla_{(\a|}R^{(1)}{}_{|\b\g)}{}^{ef}]~
-~\frac{1}{4}\nabla_{(\a|}T^{(2)}{}_{d|\b\g)} ~~~~~~~~ \eea
 \bea =
\s^{g}_{(\a\b|}\textbf{[}4i\g(\nabla_{(\a|}H^{(0)}{}_{def})\Pi^{(1)}{}_{f}{}^{ef}
-~\frac{i\g}{12}\s^{pqref}{}_{(\a\b|}A^{(1)}{}_{pqr}[\nabla_{|\g)}H^{(0)}{}_{def}]~\nn\\
-~\frac{1}{4}\nabla_{(\a|}T^{(2)}{}_{d|\b\g)}~-~2\g H^{(0)}{}_{def}
\nabla_{(\a|}R^{(1)}{}_{|\b\g)}{}^{ef} ~~~~~~~~ \eea

We need to evaluate $\nabla_{(\a|}T^{(2)}{}_{d|\b\g)}$   and
$\nabla_{(\a|}R^{(1)}{}_{|\b\g)ef}$. We will later calculate the
derivative directly but firstly we use an indirect approach. To do
this we use a first order curvature and the dimension one half
torsion at second order. This will eliminate the $chi$ terms as well
as the X tensor terms. However it also will isolate the torsion
$T^{(2)}{} _{\a\b}{}^{\la}$, and allow us to identify a candidate
for this torsion. This candidate in turn will be shown to satisfy
the dimension one half torsion, (20).

For the curvature we solve the Bianchi Identity \bea
\nabla_{(\a|}R^{(1)}{}_{|\b\g)ef}~~=~~~
T^{(0)}{}_{(\a\b|}{}{}^{~\la}R^{(1)}{}_{|\g)\la ef}
~+~T^{(1)}{}_{(\a\b|}{}{}^{~\la}R^{(0)}{}_{|\g)\la ef}
~-~T^{(0)}{}_{(\a\b|}{}^{~g}R^{(1)}{}_{|\g)gef}~\nn\\
-~T^{(1)}{}_{(\a\b|}{}{}^{~g}R^{(0)}{}_{|\g)gef}~ \eea

Using the first order constraints of Ref [4] we obtain \bea -2\g
H^{(0)}{}_{d}{}^{ef}\nabla_{(\a|}R^{(1)}{}_{|\b\g)ef}~=~
\s^{g}{}_{(\a\b|}[2\g\s_{g}{}^{\la\f}\chi_{\f}R^{(1)}{}_{|\g)\la ef}H^{(0)}{}_{d}{}^{ef}\nn\\
~+~2i\g
R^{(1)}{}_{|\g)gef}H^{(0)}{}_{d}{}^{ef}]+4\g\chi_{(\a|}R^{(1)}{}_{|\b\g)ef}H^{(0)}{}_{d}{}^{ef}
~ \eea

Similarly for the Torsion we have at second order, using equation
(20),

\bea \frac{-1}{4}\nabla_{(\a|}T^{(2)~d}{}_{|\b\g)}~=
~\frac{1}{2}\chi_{(\a|}T^{(2)}{}_{|\b\g)}{}^{d}~-~
\frac{i}{4}\s_{d(\a|\la}T^{(2)}{}_{|\b\g)}{}^{\la}~+\s^{g}{}_{(\a\b|}[
\frac{1}{4}\s_{g}{}^{\la\f}T^{(2)}{}_{|\g)\la}{}^{d}~\nn\\+~\frac{i}{4}T^{(2)}{}_{|\g)g}{}^{d}]
=~~0~~~~~~~~~ \eea Substituting these results into the derivative
$\nabla_{(\a|}H_{|\b\g)d}{}^{Order\g^{2}}$ gives the complete
expression which in turn will cancel the remaining non linear terms
in (48) exactly. We have
 \bea \frac{1}{2}\nabla_{(\a|}H_{|\b\g)d}{}^{Order 2}~=
\s^{g}_{(\a\b|}[2i\g\nabla_{|\g)}(H^{(0)}{}_{def}H^{(0)}{}_{g}{}^{ef})^{Order
2}~+~4i\g \nabla_{|\g)}H^{(0)}{}_{def}\Pi^{(1)}{}_{g}{}^{ef}]~~~\nn\\
~-~\frac{i\g}{12}
\s^{pqref}{}_{(\a\b|}[\nabla_{|\g)}(H^{(0)}{}_{def})]A^{(1)}{}_{pqr}~~
+~\frac{1}{2}\chi_{(\a|}T^{(2)}{}_{|\b\g)}{}_{d}~~~~~~~~~~~~~~~\nn\\
-~\frac{i}{4}\s_{d(\a|\la}T^{(2)}{}_{|\b\g)}{}^{\la}~
+~\s^{g}_{(\a\b|} [(\frac{1}{4})
\s_{g}{}^{\la\f}\chi_{\f}T^{(2)}{}_{|\g)\la d}
~+~\frac{i}{4}T^{(2)}{}_{|\g)gd}]~~~~~~~~~~~~~~~~~\nn\\
+~\s^{g}{}_{(\a\b|}[2\g\s_{g}{}^{\la\f}\chi_{\f}
R^{(1)}{}_{|\g)\la}{}_{ef} H^{(0)}{}_{d}{}^{ef}~+~2i\g
H^{(0)}{}_{def}R^{(1)}{}_{|\g)g}{}^{ef}]~~~~~~~~~~~~~\nn\\
+~4\g
H^{(0)}{}_{def}\chi_{(\a|}R^{(1)}{}_{|\b\g)}{}^{ef}~~~~~~~~~~~~~
\eea After substitution if the derivative term (58) into (48) we
find many cancelations and arrive at a considerably reduced $\chi$
free

\bea \frac{i}{2}\s^{g}{}_{(\a\b|}H^{(2)}{}_{g|\g)d}=~~~~
\s^{g}_{(\a\b|}\textbf{[}+2i
\g\nabla_{|\g)}(H^{(0)}{}_{def}H^{(0)}{}_{g}{}^{ef})~-~~
{2i\g}R^{(1)}{}_{|\g)[d|}{}^{ef}H^{(0)}{}_{|g]ef}~~\nn\\
-~2i\g\Pi^{(1)}{}_{g}{}^{ef}[R^{(0)}{}_{|\g)def}~-
~2\nabla_{|\g)}H^{(0)}{}_{def}]\textbf{]}~~\nn\\
+~\frac{i\g}{24}\s^{pqref}{}_{(\a\b|}A^{(1)}{}_{pqr}
[R^{(0)}{}_{|\g)d}{}_{ef}~-~2\nabla_{|\g)}H^{(0)}{}_{def}]~\nn\\
-~~\frac{i}{2}\s_{d(\a|\la}T^{(2)}{}_{|\b\g)}{}^{\la}~
+~~\s^{g}{}_{(\a\b|}\frac{i}{2}T^{(2)}{}_{d|\g) g}~~ \eea

We write the expression this way for some transparency. Now we
consider the sigma five part. Although we do not do it now, we note
that the term with $R^{(0)}{}_{\g d}{}^{ef}$ allows it to be written
as a solvable term because the identity,
 \bea
\s^{pqref}{}_{(\a\b|}\s_{e|\g)\f}=~=~-~\s^{pqref}{}_{\f(\g|}\s_{e|\a\b)}
\eea
 Hence we have using the constraints in [4],
\bea +~\frac{i\g}{24}\s^{pqref}{}_{(\a\b|}A^{(1)}{}_{pqr}
[R^{(0)}{}_{|\g)d}{}_{ef}~-~2\nabla_{|\g)}H^{(0)}{}_{def}]~=~
+~\frac{\g}{24}\s^{pqref}{}_{(\a\b|}A^{(1)}{}_{pqr}\s_{d|\g)\f}T_{ef}{}^{\f}
\eea
 After a long calculation (see appendix V) we can show that
 \bea
+~\frac{\g}{24}
\s^{pqref}{}_{(\a\b|}A^{(1)}{}_{pqr}\s_{d|\g)\f}T_{ef}{}^{\f}~~~~~~~~~~~~~~~~~~~~~\nn\\
~=-~\frac{\g}{4}\s^{g}{}_{(\a\b}A^{(1)}{}_{g}{}^{ef}\s_{d|\g)\f}T_{ef}{}^{\f}~+~4i\g^{2}
\s_{e(\a|\eps}\s_{f|\b)\tau}\s_{d|\g)\f}T_{kp}{}^{\eps}T^{kp\tau}T^{ef\f}
\eea

It was the second term in RHS of the above expression, (62), that
caused the problem of closure. It is not in a solvable form, and it
cannot be written as such. We could avoid the problem by absorbing
the second term of the above equation into the torsion
$T^{(2)}{}^{\la}{}_{\b\g}$. The terms in (62) all have a similar
$\s_{d\g \f}$ coefficient and also similar to that of
$T^{(2)}{}^{\la}{}_{\b\g}$. It remains to be considered as to what
combination should be absorbed into $T^{(2)}{}^{\la}{}_{\b\g}$. We
could have
 \bea \frac{i}{2}\s_{d(\a|\la}T^{(2)}{}_{|\b\g)}{}^{\la}
~=~+~4i\g^{2}
\s_{e(\a|\eps}\s_{f|\b)\f}\s_{d|\g)\la}T_{kp}{}^{\eps}T^{kp\f}T^{ef\la}
\eea
 However we have evidence from the dimension one half torsion, (20), that we should
 in fact choose the whole quantity with that particular sigma matrix structure as follows
 \bea
i\s_{d(\a|\la}T^{(2)}{}_{|\b\g)}{}^{\la} =+~\frac{i\g}{12}
\s^{pqref}{}_{(\a\b|}A^{(1)}{}_{pqr}[R^{(0)}{}_{|\g)d}{}_{ef}~-~2\nabla_{|\g)}H^{(0)}{}_{def}]\nn\\
\eea Or
 \bea
T^{(2)}{}_{\a\b}{}^{\la}~=~-~\frac{i\g}{12}\s^{pqref}{}_{\a\b}A^{(1)}{}_{pqr}T_{ef}{}^{\f}
\eea

This scenario will give for $H^{(2)}{}_{g \g d}$,

\bea \frac{i}{2}\s^{g}{}_{(\a\b|}H^{(2)}{}_{g|\g)d}=~~~~
\s^{g}{}_{(\a\b|}\textbf{[}+2i
\g\nabla_{|\g)}(H^{(0)}{}_{def}H^{(0)}{}_{g}{}^{ef})~-~~
{2i\g}R^{(1)}{}_{|\g)[d|}{}^{ef}H^{(0)}{}_{|g]ef}~~\nn\\
-~2i\g [\Pi^{(1)}{}_{g}{}^{ef}][R^{(0)}{}_{|\g)def}~-
~2\nabla_{|\g)}H^{(0)}{}_{def}]\textbf{]}~~\nn\\
+~~\s^{g}{}_{(\a\b|}\frac{i}{2}T^{(2)}{}_{d|\g) g}~~ \eea

This is written in terms of the as yet unknown torsion $T^{(2)}{}_{d
\g g}$.

\section{Closure Using the X Tensor}

We now propose an Anzatz for the X tensor and show that in
conjunction with the results (64), and (39), that we can indeed
close the H sector by a different route. We later show that these
results close the dimension one half torsion.

We have, using the dimension one half torsion (20), \bea
-~~\frac{i}{2}\s_{d(\a|\la}T^{(2)}{}_{|\b\g)}{}^{\la}~
+~~\s^{g}{}_{(\a\b|}\frac{i}{2}T^{(2)}{}_{d|\g) g}~~\nn\\
~=~\frac{1}{2}T^{(0)}{}_{(\a\b|}{}^{\la}T^{(2)}{}_{|\g)\la}{}^{d}~
-~\frac{1}{2}\nabla_{(\a|}T^{(2)}{}_{|\b\g)}{}^{d} \eea The last two
terms in equation (59) also appear as in this combination.

We now say let \bea
T^{(2)}{}_{\a\b}{}^{d}=\s^{pqref}{}_{\a\b}[X_{pqrefd}+Y_{pqrefd}]\eea
where $X_{pqrefd}-~\frac{i\g}{6}\s^{pqref}{}_{\a\b}$.  We find that
$Y_{pqrefd}=0$
 is sufficient to close the H sector and torsion sector identities.
 Hence we
 choose the Ansatz
 \bea
T^{(2)}{}_{\a\b}{}^{d}~=-~\frac{i\g}{6}\s^{pqref}{}_{\a\b}
H^{(0)}{}^{d}{}_{ef}A^{(1)}{}_{pqr}\eea

Therefore

\bea -~~\frac{i}{2}\s_{d(\a|\la}T^{(2)}_{|\b\g)}{}^{\la}~
+~~\s^{g}{}_{(\a\b|}\frac{i}{2}T^{(2)}{}_{d|\g) g}~~~~~~~~~~~~~~~~~~~~~~~~~~~~~~~\nn\\
~=~\frac{1}{2}T^{(0)}{}_{(\a\b|}{}^{\la}[-~\frac{i\g}{6}\s^{pqref}{}_{|\g)\la}
H^{(0)}{}_{def}A^{(1)}{}_{pqr} ]~
-~\frac{1}{2}\nabla_{(\a|}[-~\frac{i\g}{6}\s^{pqref}{}_{|\b\g)}
H^{(0)}{}_{def}A^{(1)}{}_{pqr} ]\eea

 \bea=~\frac{1}{2}T^{(0)}{}_{(\a\b|}{}^{\la}[-~\frac{i\g}{6}\s^{pqref}{}_{|\g)\la}
H^{(0)}{}_{def}A^{(1)}{}_{pqr} ]~
-~\frac{1}{2}[-~\frac{i\g}{6}\s^{pqref}{}_{|\b\g)}
H^{(0)}{}_{def}[\nabla_{(\a|}A^{(1)}{}_{pqr}]\nn\\
-~\frac{1}{2}[-~\frac{i\g}{6}\s^{pqref}{}_{|\b\g)}
[\nabla_{(\a|}H^{(0)}{}_{def}]A^{(1)}{}_{pqr} \eea

We now use equation (39) of our recipe to get

 \bea -~\frac{i}{2}\s_{d(\a|\la}T^{(2)}{}_{|\b\g)}{}^{\la}~
+~\s^{g}{}_{(\a\b|}\frac{i}{2}T^{(2)}{}_{d|\g) g}~
~=-\frac{i\g}{2}\s^{g}{}_{(\a\b|}[\nabla_{|\g)}A^{(1)}{}_{gef}]H^{(0)}{}_{d}{}^{ef}\nn\\
+2i\g\s^{g}{}_{(\a\b|}R^{(1)}{}_{|\g)gef}H^{(0)}{}_{d}{}^{ef}
+~\frac{i\g}{12}[\s^{pqref}{}_{(\a\b|}
[\nabla_{|\g)}H^{(0)}{}_{def}]A^{(1)}{}_{pqr} \eea

Incorporating the result (72) into (59) gives

\bea~\frac{i}{2}\s^{g}{}_{(\a\b|}H^{(2)}{}_{g|\g)d}=~
\s^{g}_{(\a\b|}\textbf{[}+2i
\g\nabla_{|\g)}(H^{(0)}{}_{def}H^{(0)}{}_{g}{}^{ef})~-~~
{2i\g}R^{(1)}{}_{|\g)[d|}{}^{ef}H^{(0)}{}_{|g]ef}~~\nn\\
-~2i\g[\Pi^{(1)}{}_{g}{}^{ef}][R^{(0)}{}_{|\g)def}~-
~2\nabla_{|\g)}H^{(0)}{}_{def}]\textbf{]}~~\nn\\
+~\frac{i\g}{24}\s^{pqref}{}_{(\a\b|}A^{(1)}{}_{pqr}
[R^{(0)}{}_{|\g)d}{}_{ef}]\nn\\
~-~\frac{i\g}{2}\s^{g}{}_{(\a\b|}\nabla_{|\g)}A^{(1)}{}_{gef}H^{(0)}{}_{d}{}^{ef}
+2i\g \s^{g}{}_{(\a\b|}R^{(1)}{}_{|\g)gef}H^{(0)}{}_{d}{}^{ef}\nn\\
~~\eea

Furthermore using  equation (41) we have

\bea \frac{i\g}{12}\s^{pqref}{}_{(\a\b|}A^{(1)}{}_{pqr}
R^{(0)}{}_{\g)}{}_{def}=~+\frac{\g}{6}\s^{g}{}_{(\a\b|}\s^{pqre}{}_{g}{}_{|\g)\f}A^{(1)}{}_{pqr}T^{(0)}{}_{d
e}^{}{\f} \eea Hence we finally obtain a fully solvable form as
follows

\bea~\frac{i}{2}\s^{g}{}_{(\a\b|}H^{(2)}{}_{g|\g)d}=\s^{g}_{(\a\b|}\textbf{[}+2i
\g\nabla_{|\g)}(H^{(0)}{}_{def}H^{(0)}{}_{g}{}^{ef})-
{2i\g}R^{(1)}{}_{|\g) d|}{}^{ef}H^{(0)}{}_{g ef}~\nn\\
-2i\g[\Pi^{(1)}{}_{g}{}^{ef}][R^{(0)}{}_{|\g)def}-
2\nabla_{|\g)}H^{(0)}{}_{def}]\textbf{]}\nn\\
+\frac{\g}{12}\s^{g}{}_{(\a\b|}\s^{pqre}{}_{g}{}_{|\g)\f}A^{(1)}{}_{pqr}T^{(0)}{}_{d
e}^{}{\f}-\frac{i\g}{2}\s^{g}{}_{(\a\b|}\nabla_{|\g)}A^{(1)}{}_{gef}H^{(0)}{}_{d}{}^{ef}\nn\\
\eea Hence we have solved the first part old problem. Also the
residual terms in (75), should be those obtained in the solution to
the dimension one half torsion. In fact in the next section we find
that that is just so, hence adding good support to our Ansatz for
$T^{(2)}{}_{\b\g}{}^{g}$ and result for the torsion
$T^{(2)}{}_{\b\g}{}^{\la}$. We show that this is true in this
scenario, and we also show that it is true in the case where we take
the direct derivative $H^{(2)}{}_{\a\b d}$ without using the
curvature or torsion to eliminate non-linear terms.
\section{Dimension One Half Torsion}
 We now look at  the dimension one
 half torsion, (equation (20)). At second order since all relevant first order quantities are zero
this becomes
 \bea T^{(0)}{}_{(\a\b|}{}^{\la}T^{(2)}{}_{|\g)\la}{}^{d}~
+~T^{(2)}{}_{(\a\b|}{}^{\la}T^{(0)}{}_{|\g)\la}{}^{d}
~-~T^{(0)}{}_{(\a\b|}{}^{g}T^{(2)}{}_{|\g)g}{}^{d}~
-~\nabla_{(\a|}T^{(2)}{}_{|\b\g)}{}^{d}~=~0 \eea We have the
candidate for the X tensor, equation (69). We also had the candidate
for the complete term , $i\s_{d(\a|\la}T^{(2)}{}_{|\b\g)}{}^{\la}$,
(64). Hence substitution of these results into the torsion (76)
gives

\bea
T^{(0)}{}_{(\a\b|}{}^{\la}\s^{pqref}{}_{|\g)\la}A^{(1)}{}_{pqr}H^{(0)}{}_{def}[-\frac{i\g}{6}]
+~\frac{i\g}{12}\s^{pqref}{}_{(\a\b|}A^{(1)}{}_{pqr}
R^{(0)}{}_{\g)}{}_{def}\nn\\
~-~\frac{i\g}{6}\s^{pqref}{}_{(\a\b|}A^{(1)}{}_{pqr}\nabla_{|\g)}H^{(0)}{}_{def}
~+~\frac{i\g}{6}\s^{pqref}{}_{(\a\b|}A^{(1)}{}_{pqr}\nabla_{|\g)}H^{(0)}{}_{def}\nn\\
+~\frac{i\g}{6}\s^{pqref}{}_{(\a\b|}H^{(0)}{}_{def}\nabla_{|\g)}A^{(1)}{}_{pqr}\mid_{\chi}
+~\frac{i\g}{6}\s^{pqref}{}_{(\a\b|}H^{(0)}{}_{def}\nabla_{|\g)}A^{(1)}{}_{pqr}\mid_{R}\nn\\
-i\s^{g}{}_{(\a\b|}T^{(2)}{}_{|\g)}{}_{gd} =0\eea

Again using (39) and (41) gives many cancelations of terms which
otherwise would be intractable. Hence we obtain the very short
result,

\bea +~i\s^{g}{}_{(\a\b|}T^{(2)}{}_{|\g) gd}~=
-i\g\s^{g}{}_{(\a\b|}\nabla_{|\g)}A^{(1)}{}_{gef}H^{(0)}{}_{d}{}^{ef}\nn\\
4i\g\s^{g}{}_{(\a\b|}R^{(1)}{}_{|\g)gef}H^{(0)}{}_{def}
~+~\frac{\g}{6}\s^{g}{}_{(\a\b|}\s^{pqre}{}_{g}{}_{|\g)\f}A^{(1)}{}_{pqr}T^{(0)}{}_{de}{}^{\f}
\eea
 This is in agreement with comparing (59) to (75).

\section{Direct Derivative Check}

We now look at an extremely interesting check on our work. It will
be based on a previously unnoticed observation. We began by
calculating $H^{(2)}{}_{\b\g d}$. We found this quite easily. It
contained the X tensor. We needed the spinor derivative of
$H^{(2)}{}_{\b\g d}$ to find $H^{(2)}{}_{\a b g}$. Since we did not
know the form or the X tensor we eliminated it as well as all $\chi$
terms by using the dimension one half torsion and also a curvature
to replace those terms. Hence we did not calculate this derivative
directly. We found a solution for $H^{(2)}{}_{\a b g}$ which
contained $T^{(2)}{}_{\a b}{}^{g}$ as an unknown. In the process we
also identified $T^{(2)}{}_{\b\g}{}^{\la}$. We then proposed a
candidate for the X tensor and showed that it could successfully
close the second H sector Bianchi Identities and the dimension one
half torsion. This candidate also closed the dimension one half
torsion, and produced a result for $T^{(2)}{}_{\a b}{}^{g}$ . We
also found the exact same result for $T^{(2)}{}_{\a b}{}^{g}$ by
comparing the two results for $H^{(2)}{}_{\a b g}$.

In the following we show agreement and consistency with the results
of the solution to the dimension one half torsion, and our results
in the H sector, while taking the direct derivative of
$H^{(2)}{}_{\a \b g}$.

We make the very convenient observation that the following
quantities are interchangeable. We find the following useful result
which allows us to make a comparison between the direct method and
the method used in  part (5) that employed a torsion and curvature
to eliminate non linear terms. We note conveniently that

\bea \nabla_{\a}L^{(1)}{}_{abc}
=i\g\s_{[g|\g\f}T^{(0)}{}_{kl}{}^{\f}R^{kl}{}_{|ef]}\eea

 Using the result $\s^{g}{}_{(\a\b|}\s_{g |\g)\la}=0 $ along
 with the result for $R^{(1)}{}_{\g}{}_{abc}$ as given in \cite{4}, we can write

 \bea \nabla_{\g}L^{(1)}{}_{abc}= R^{(1)}{}_{\g}{}_{abc}\eea

This is a crucial observation which will have other roles. We note
here that the result for the same derivative in \cite{2} will
 fail to work in the following closure, whereas that in \cite{4} will indeed work, and
 so the coefficient in \cite{4} seems to be correct.

We found the H sector solution,  equation (59).  We can write it as
follows

 \bea \frac{i}{2}\s^{g}{}_{(\a\b|}H^{(2)}{}_{g|\g)d}=~ \s^{g}{}_{(\a\b|}\textbf{[}+2i
\g\nabla_{|\g)}(H^{(0)}{}_{def}H^{(0)}{}_{g}{}^{ef})~-~
{2i\g}R^{(1)}{}_{|\g)d}{~}^{ef}H^{(0)}{}_{gef}\nn\\~+~{2i\g}R^{(1)}{}_{|\g)g}{}^{ef}H^{(0)}{}_{def}~
-2i\g \Pi^{(1)}{}_{g}{}^{ef}R^{(0)}{}_{|\g)def} +4i\g
L^{(1)}{}_{g}{}^{ef}\nabla_{|\g)}H^{(0)}{}_{def}~\nn\\
-\frac{ i\g}{2}A^{(1)}{}_{g}{}^{ef}\nabla_{|\g)}H^{(0)}{}_{def}
+~~\s^{g}{}_{(\a\b|}\frac{i}{2}T^{(2)}{}_{d|\g) g}~~ \eea

Also we found

\bea +~\frac{i}{2}\s^{g}{}_{(\a\b|}T^{(2)}{}_{|\g) gd}~=~
-\frac{i\g}{2}\s^{g}{}_{(\a\b|}\nabla_{|\g)}A^{(1)}{}_{gef}H^{(0)}{}_{d}{}^{ef}\nn\\
+~2i\g\s^{g}{}_{(\a\b|} R^{(1)}{}_{|\g) gef}H^{(0)}{}_{d}{}^{ef}
~+~\frac{\g}{12}\s^{g}{}_{(\a\b|}\s^{pqre}{}_{g}{}_{|\g)\f}A^{(1)}{}_{pqr}T^{(0)}{}_{de}{}^{\f}
\eea

Hence we have

 \bea  \frac{i}{2}\s^{g}{}_{(\a\b|}H^{(2)}{}_{g|\g)d}=~\s^{g}_{(\a\b|}\textbf{[}+2i
\g\nabla_{|\g)}(H^{(0)}{}_{def}H^{(0)}{}_{g}{}^{ef})~\nn\\
-~{2i\g}R^{(1)}{}_{|\g)d}{~}^{ef}H^{(0)}{}_{gef}~+~{4i\g}R^{(1)}{}_{|\g)g}{~}^{ef}H^{(0)}{}_{def}~\nn\\
-2i\g L^{(1)}{}_{g}{}^{ef}R^{(0)}{}_{|\g)def}
+~\frac{i\g}{4}A^{(1)}{}_{g}{}^{ef}R^{(0)}{}_{|\g)d}{}^{ef}\nn\\
+4i\g L^{(1)}{}_{g}{}^{ef}\nabla_{|\g)}H^{(0)}{}_{def}~
-\frac{i\g}{2}A^{(1)}{}_{g}{}^{ef}\nabla_{|\g)}H^{(0)}{}_{def}\nn\\
-\frac{i\g}{2}\nabla_{|\g)}A^{(1)}{}_{gef}H^{(0)}{}_{d}{}^{ef}
~+~\frac{\g}{12}\s^{pqre}{}_{g}{}_{|\g)\f}A^{(1)}{}_{pqr}T^{(0)}{}_{de}{}^{\f}]
\eea

In terms of the X tensor we obtained

\bea -~\frac{i}{4}\s_{d(\a|\la}T^{(2)}_{|\b\g)}{}^{\la}~
+~~\s^{g}{}_{(\a\b|}\frac{i}{4}T^{(2)}{}_{d|\g) g}~
~=~\frac{1}{4}T^{(0)}{}_{(\a\b|}{}^{\la}[-~\frac{i\g}{6}\s^{pqref}{}_{|\g)\la}
H^{(0)}{}_{def}A^{(1)}{}_{pqr} ]~\nn\\
-~\frac{1}{4}\nabla_{(\a|}[-~\frac{i\g}{6}\s^{pqref}{}_{|\b\g)}
H^{(0)}{}_{def}A^{(1)}{}_{pqr} ]\eea

Working out (47) term by term by substituting in the first order
constraints as before and recalling that $\s^{g}{}_{(\a\b|}\s_{g
|\g)\f}=0$ eliminates one term, we find after a lengthy calculation
that

\bea \frac{+i}{2}\s^{g}{}_{(\a\b|}H^{(2)}{}_{g|\g)d}~=~
{\frac{1}{2}}\nabla_{(\a|}H_{|\b\g)d}{}^{Order\g^{2}}~
~-~\frac{i}{4}\s_{d(\a|\la}T^{(2)}{}_{|\a\b)}{}^{\la}~+~\frac{i}{4}\s^{g}_{(\a\b|}T^{(2)}{}_{d|\g)g}\nn\\
+~T^{(0)}_{(\a\b|}{}^{\la}\s^{pqr}{}_{ef}{}_{|\g)\la}[+~\frac{i\g}{12}H^{(0)}{}_{def}A^{(1)}{}_{pqr}~+~\frac{1}{4}X_{pqrefd}]
~+~\s^{g}{}_{(\a\b|}[{-2i\g}H^{(0)}{}_{gef}R^{(1)}{}_{|\g)d}{}^{ef}~\nn\\
{-2i\g}\Pi^{(1)}{}_{gef}R^{(0)}{}_{|\g)d}{}^{ef}]
+~\frac{i\g}{24}\s^{pqr}{}_{ef}{}_{(\a\b|}R^{(0)}{}_{|\g)d}{}^{ef}A^{(1)}{}_{pqr}\nn\\
\eea

We need $\frac{1}{2}\nabla_{(\a|}H_{|\b\g)d}{}^{Order\g^{2}}$. We
take the derivative of (46) directly. The derivative is \bea
\frac{1}{2}\nabla_{(\a|}H_{|\b\g)d}{}^{Order 2}~~=~~~
\s^{g}_{(\a\b|}[2i\g\nabla_{|\g)}(H^{(0)}{}_{def}H^{(0)}{}_{g}{}^{ef})^{Order (2))}~\nn\\
+~4i\g\nabla_{|\g)}[H^{(0)}{}_{def}\Pi^{(1)}{}_{g}{}^{ef}]~~
+\frac{1}{2}\s^{pqref}{}_{(\a\b|}\nabla_{|\g)}[(-\frac{i\g}{6}H^{(0)}{}_{def})A^{(1)}{}_{pqr}~\nn\\
+~\frac{1}{2}X_{pqref}]~\eea
~~~~~~\nn\\
The whole thing becomes

 \bea
\frac{+i}{2}\s^{g}{}_{(\a\b|}H^{(2)}{}_{g|\g)d}~=
~-~\frac{i}{4}\s_{d(\a|\la}T^{(2)}{}_{|\a\b)}{}^{\la}~+~\frac{i}{4}\s^{g}_{(\a\b|}T^{(2)}{}_{d|\g)g}\nn\\
~+~\s^{g}_{(\a\b|}[2i\g\nabla_{|\g)}(H^{(0)}{}_{def}H^{(0)}{}_{g}{}^{ef})^{Order
(2)}~
+~4i\g\nabla_{|\g)}[H^{(0)}{}_{def}\Pi^{(1)}{}_{g}{}^{ef}]]~~\nn\\
+\frac{1}{2}\s^{pqref}{}_{(\a\b|}\nabla_{|\g)}[(-\frac{i\g}{6}H^{(0)}{}_{def})A^{(1)}{}_{pqr}~
+~\frac{1}{2}X_{pqref}]\nn\\
+~T^{(0)}_{(\a\b|}{}^{\la}\s^{pqref}{}_{|\g)\la}[+~\frac{i\g}{12}H^{(0)}{}_{def}A^{(1)}{}_{pqr}~+~\frac{1}{4}X_{pqrefd}]
~+~\s^{g}{}_{(\a\b|}[{-2i\g}H^{(0)}{}_{gef}R^{(1)}{}_{|\g)d}{}^{ef}~\nn\\
{-2i\g}\Pi^{(1)}{}_{gef}R^{(0)}{}_{|\g)d}{}^{ef}]
+~\frac{i\g}{24}\s^{pqr}{}_{ef}{}_{(\a\b|}R^{(0)}{}_{|\g)d}{}^{ef}A^{(1)}{}_{pqr}
~~~~~~~~~~~~~~~~~~~~~~\eea

Substituting in for the X tensor and gives

 \bea
\frac{+i}{2}\s^{g}{}_{(\a\b|}H^{(2)}{}_{g|\g)d}~=
~\frac{1}{4}T^{(0)}{}_{(\a\b|}{}^{\la}[-~\frac{i\g}{6}\s^{pqref}{}_{|\g)\la}
H^{(0)}{}_{def}A^{(1)}{}_{pqr} ]~\nn\\
-~\frac{1}{4}\nabla_{(\a|}[-~\frac{i\g}{6}\s^{pqref}{}_{|\b\g)}
H^{(0)}{}_{def}A^{(1)}{}_{pqr} ]
~+~\s^{g}_{(\a\b|}[2i\g\nabla_{|\g)}(H^{(0)}{}_{def}H^{(0)}{}_{g}{}^{ef})^{Order
(2)}~\nn\\
+~4i\g\nabla_{|\g)}[H^{(0)}{}_{def}\Pi^{(1)}{}_{g}{}^{ef}]~
+\s^{pqref}{}_{(\a\b|}\nabla_{|\g)}[(-\frac{i\g}{24}H^{(0)}{}_{def})A^{(1)}{}_{pqr}]\nn\\
+~T^{(0)}_{(\a\b|}{}^{\la}\s^{pqref}{}_{|\g)\la}[+~\frac{i\g}{24}H^{(0)}{}_{def}A^{(1)}{}_{pqr}]
~+~\s^{g}{}_{(\a\b|}[{-2i\g}H^{(0)}{}_{gef}R^{(1)}{}_{|\g)d}{}^{ef}~\nn\\
{-2i\g}\Pi^{(1)}{}_{gef}R^{(0)}{}_{|\g)d}{}^{ef}]
+~\frac{i\g}{24}\s^{pqr}{}_{ef}{}_{(\a\b|}R^{(0)}{}_{|\g)d}{}^{ef}A^{(1)}{}_{pqr}
~~~~~~~~~~~~~~~~~~~~~~\eea

Terms neatly cancel to get

 \bea
\frac{+i}{2}\s^{g}{}_{(\a\b|}H^{(2)}{}_{g|\g)d}~=
~+~\s^{g}_{(\a\b|}[2i\g\nabla_{|\g)}(H^{(0)}{}_{def}H^{(0)}{}_{g}{}^{ef})^{Order(2)}~\nn\\
+~4i\g\nabla_{|\g)}[H^{(0)}{}_{def}\Pi^{(1)}{}_{g}{}^{ef}]~-~2i\g H^{(0)}{}_{gef}R^{(1)}{}_{|\g)d}{}^{ef}~\nn\\
{-2i\g}\Pi^{(1)}{}_{gef}R^{(0)}{}_{|\g)d}{}^{ef}]
+~\frac{i\g}{24}\s^{pqr}{}_{ef}{}_{(\a\b|}R^{(0)}{}_{|\g)d}{}^{ef}A^{(1)}{}_{pqr}
~\eea

Using \bea
\s^{pqref}{}_{(\a\b|}\s_{e|\g)\f}=~=~-~\s^{pqref}{}_{(\g|}{}_{\f}\s_{e|\a\b)}
\eea

Gives the soluble form
 \bea
\frac{+i}{2}\s^{g}{}_{(\a\b|}H^{(2)}_{g|\g)d}~=
~+~\s^{g}_{(\a\b|}[2i\g\nabla_{|\g)}(H^{(0)}{}_{def}H^{(0)}{}_{g}{}^{ef}){}^{Order(2)}~\nn\\
+~4i\g\nabla_{|\g)}[H^{(0)}{}_{def}\Pi^{(1)}{}_{g}{}^{ef}]~-~2i\g H^{(0)}{}_{gef}R^{(1)}{}_{|\g)d}{}^{ef}~\nn\\
{-2i\g}\Pi^{(1)}{}_{gef}R^{(0)}{}_{|\g)d}{}^{ef}]
+~\frac{\g}{12}\s^{g}{}_{(\a\b|}\s^{pqre}{}_{g}{}_{|g)\f}T^{(0)}{}_{de}{}^{f}A^{(1)}{}_{pqr}
~\eea

 \bea
\Rightarrow \frac{+i}{2}\s^{g}{}_{(\a\b|}H^{(2)}{}_{g|\g)d}~=
~+~\s^{g}_{(\a\b|}[2i\g\nabla_{|\g)}(H^{(0)}{}_{def}H^{(0)}{}_{g}{}^{ef})^{Order(2)}~
+~4i\g\nabla_{|\g)}[H^{(0)}{}_{def}]\Pi^{(1)}{}_{g}{}^{ef}\nn\\
+~4i\g H^{(0)}{}_{def}\nabla_{|\g)}[\Pi^{(1)}{}_{g}{}^{ef}] ~-~2i\g
H^{(0)}{}_{gef}R^{(1)}{}_{|\g)d}{}^{ef}
{-2i\g}\Pi^{(1)}{}_{gef}R^{(0)}{}_{|\g)d}{}^{ef}]\nn\\
+~\frac{\g}{12}\s^{g}{}_{(\a\b|}\s^{pqre}{}_{g}{}_{\g)\f}T^{(0)}{}_{de}{}^{\f}A^{(1)}{}_{pqr}\nn\\~
~\eea

The two results (66) and (92) at first sight do not seem to
coincide.
 We need to work out the $\Pi$ derivative in equation (92) to enable a comparison.  For this our crucial
observation is given in (80). We have therefore

\bea ~4i\g
\s^{g}{}_{(\a\b|}H^{(0)}{}_{def}\nabla_{|\g)}[\Pi^{(1)}{}_{g}{}^{ef}]~=~
 ~4i\g \s^{g}{}_{(\a\b|}H^{(0)}{}_{def}\nabla_{|\g)}[L^{(1)}{}_{g}{}^{ef}~-~\frac{1}{8}A^{(1)}{}_{g}{}^{ef}]\nn\\
 =~~+4i\g \s^{g}{}_{(\a\b|}H^{(0)}{}_{d}{}^{ef}R^{(1)}{}_{|\g) gef}
 ~-~\frac{i\g}{2}\s^{g}{}_{(\a\b|}H^{(0)}{}_{d}{}^{ef}\nabla_{|\g)}A^{(1)}{}_{gef}\nn\\\eea

 \bea
\Rightarrow \frac{+i}{2}\s^{g}{}_{(\a\b|}H^{(2)}{}_{g|\g)d}~=
~+~\s^{g}_{(\a\b|}[2i\g\nabla_{|\g)}(H^{(0)}{}_{def}H^{(0)}{}_{g}{}^{ef})^{Order(2)}~
+~4i\g(\nabla_{|\g)}H^{(0)}{}_{def})\Pi^{(1)}{}_{g}{}^{ef}\nn\\
+4i\g H^{(0)}{}_{d}{}^{ef}R^{(1)}{}_{|\g) gef}
 ~-~\frac{i\g}{2}H^{(0)}{}_{d}{}^{ef}\nabla_{|\g)}A^{(1)}{}_{gef}]\nn\\~
-~2i\g H^{(0)}{}_{gef}R^{(1)}{}_{|\g)d}{}^{ef}
{-2i\g}\Pi^{(1)}{}_{gef}R^{(0)}{}_{|\g)d}{}^{ef}]
+~\frac{\g}{12}\s^{g}{}_{(\a\b|}\s^{pqre}{}_{g}{}_{|\g)\f}T^{(0)}{}_{de}{}^{\f}A^{(1)}{}_{pqr}\nn\\
~\eea

\bea~~=~\s^{g}_{(\a\b|}\textbf{[}+2i
\g\nabla_{|\g)}(H^{(0)}{}_{def}H^{(0)}{}_{g}{}^{ef})~-~~
{2i\g}R^{(1)}{}_{|\g) [d|}{~}^{ef}H^{(0)}{}_{|g] ef}~~\nn\\
-~2i\g[\Pi^{(1)}{}_{g}{}^{ef}][R^{(0)}{}_{|\g)d}{}^{ef}~-
~2\nabla_{|\g)}H^{(0)}{}_{def}]\textbf{]}~~\nn\\
+\frac{\g}{12}\s^{g}{}_{(\a\b|}\s^{pqre}{}_{g}{}_{|\g)\f}A^{(1)}{}_{pqr}T^{(0)}{}_{d
e}^{}{\f}~-~\frac{i\g}{2}\s^{g}{}_{(\a\b|}\nabla_{|\g)}A^{(1)}{}_{gef}H^{(0)}{}_{d}{}^{ef}
\eea

Hence we find exact agreement with between (66) and (95), using also
(78).

\section{Torsion for $T^{(2)}{}_{\a d}{}^{\d}$ at Dimension One}

This is based on the constraints listed in [2]. We have the Bianchi
identity at dimension one as follows,

\bea
T_{(\a\b|}{}^{\la}T_{|\g)\la}{}^{\d}~-~T_{(\a\b|}{}^{g}T_{|\g)g}{}^{\d}
~-~\nabla_{(\a|}T_{|\b\g)}{}^{\d} -~\frac{1}{4}R_{(\a\b|
de}\s^{de}{}_{|\g)}{}^{\d} ~=~0 \eea

At second order it becomes

\bea T^{(0)}{}_{(\a\b|}{}^{\la}T^{(2)}{}_{|\g)\la}{}^{\d}~+
~T^{(2)}{}_{(\a\b|}{}^{\la}T^{(0)}{}_{|\g)\la}{}^{\d}
-i\s^{g}{}_{(\a\b|}T^{(2)}{}_{|\g)g}{}^{\d}~-~\nabla_{(\a|}T^{(2)}{}_{|\b\g)}{}^{\d}\nn\\
-~\nabla_{(\a|}T^{(0)}{}_{|\b\g)}{}^{\d}{}^{(Order\g^{2})}-~\frac{1}{4}R^{(2)}{}_{(\a\b|
de}\s^{de}{}_{|\g)}{}^{\d} ~=~0\nn\\\eea

We must take care not to neglect second order contributions from the
derivative $\nabla_{\a}T^{(0)}{}_{\b\g}{}^{\d}$.

We have
\bea-~\nabla_{(\a|}T^{(0)}{}_{|\b\g)}{}^{\d}{}^{(Order\g^{2})}=
[2\d_{(\a|}{}^{\d}\d_{|\b)}{}^{\la}+\s^{g}{}_{(a\b|}\s_{g}{}^{\d
\la}]\nabla_{|\g)}\chi_{\la}\eea

Using the form of the derivative $\nabla_{\a}\chi_{\b}$ as quoted in
reference [4] and the result that

\bea [2\d_{(\a|}^{\d}\d_{|\b)}{}^{\la}+\s^{g}{}_{(a\b|}\s_{g}{}^{\d
\la}]\s_{b|\g)\d}=0 \eea Gives \bea
-\nabla_{(\a|}T^{(0)}{}^{Order\g^{2}}{}_{|\b\g)}{}^{\d}=
[2\d_{(\a|}{}^{\d}\d_{|\b)}{}^{\la}+\s^{g}{}_{(\a\b|}\s_{g}{}^{\d
\la}]\nabla_{|\g)}\chi_{\la}=\nn\\
-\frac{i}{2}\s^{g}{}_{(\a\b|}\s^{mn}{}_{|\g)}{}^{\d}[L^{(2)}{}_{gmn}+\frac{1}{4}A^{(2)}{}_{gmn}]
\eea

We now have as before which we re-list for convenience,
 \bea
T^{(2)}{}_{\a\b}{}^{\la}~=~-\frac{i\g}{12}\s^{pqref}{}_{\a\b}A^{(1)}{}_{pqr}T^{(0)}{}_{ef}{}^{\la}
\eea Hence the torsion becomes

\bea
T^{(0)}{}_{(\a\b|}{}^{\la}[-\frac{i\g}{12}\s^{pqref}{}_{|\g)\la}A^{(1)}{}_{pqr}T^{(0)}{}_{ef}{}^{\d}]
~-~\frac{i\g}{12}\s^{pqref}{}_{(\a\b|}A^{(1)}{}_{pqr}T^{(0)}{}_{ef}{}^{\d}T^{(0)}{}_{|\g)\la}{}^{\d}
-i\s^{g}{}_{(\a\b|}T^{(2)}{}_{|\g)g}{}^{\d}\nn\\
+~\frac{i\g}{12}\s^{pqref}{}_{(\a\b|}(\nabla_{|\g)}A^{(1)}{}_{pqr})T^{(0)}{}_{ef}{}^{\d}
+~\frac{i\g}{12}\s^{pqref}{}_{(\a\b|}A^{(1)}{}_{pqr}(\nabla_{|\g)}T^{(0)}{}_{ef}{}^{\d})
-~\frac{1}{4}R^{(2)}{}_{(\a\b|de}\s^{de}{}_{|\g)}{}^{\d}~\nn\\
-\frac{i}{2}\s^{g}{}_{(\a\b|}\s^{mn}{}_{|\g)}{}^{\d}[L^{(2)}{}_{gmn}+\frac{1}{4}A^{(2)}{}_{gmn}]\nn\\
=0\nn\\ ~\eea

Using  (39) and (41) again gives gives

\bea
-\frac{i\g}{2}\s^{g}{}_{(\a\b|}[\nabla_{|\g)}A^{(1)}{}_{g}{}^{ef}]T^{(0)}{}_{ef}{}^{\d}
+2i\g\s^{g}{}_{(\a\b|}R^{(1)}{}_{|\g) g}{}^{ef}T^{(0)}{}_{ef}{}^{\d}\nn\\
~-~\frac{i\g}{12}\s^{pqref}{}_{(\a\b|}A^{(1)}{}_{pqr}T^{(0)}{}_{ef}{}^{\d}T^{(0)}{}_{|\g)\la}{}^{\d}
-i\s^{g}{}_{(\a\b|}T^{(2)}{}_{|\g)g}{}^{\d}\nn\\
+~\frac{i\g}{12}\s^{pqref}{}_{(\a\b|}A^{(1)}{}_{pqr}[-\frac{1}{4}\s^{mn}{}_{|\g)}{}^{\d}R_{efmn}
~+~T^{(0)}{}_{ef}{}^{\la}T^{(0)}{}_{|\g)\la}{}^{\d}]-
~\frac{1}{4}R^{(2)}{}_{(\a\b|de}\s^{de}{}_{|\g)}{}^{\d}~\nn\\
-\frac{i}{2}\s^{g}{}_{(\a\b|}\s^{mn}{}_{|\g)}{}^{\d}[L^{(2)}{}_{gmn}+\frac{1}{4}A^{(2)}{}_{gmn}]=0
~~~\eea

The use of the derivative (35) allows a cancelation of what would
otherwise have been an insolvable term. We find

\bea
-\frac{i\g}{2}\s^{g}{}_{(\a\b|}[\nabla_{|\g)}A^{(1)}{}_{g}{}^{ef}]T^{(0)}{}_{ef}{}^{\d}
+2i\g\s^{g}{}_{(\a\b|}R^{(1)}{}_{|\g)
g}{}^{ef}T^{(0)}{}_{ef}{}^{\d}\nn\\
-i\s^{g}{}_{(\a\b|}T^{(2)}{}_{|\g)g}{}^{\d}
+~\frac{i\g}{12}\s^{pqref}{}_{(\a\b|}A^{(1)}{}_{pqr}[-\frac{1}{4}\s^{mn}{}_{|\g)}{}^{\d}R_{efmn}]
-~\frac{1}{4}R^{(2)}{}_{(\a\b|mn}\s^{mn}{}_{|\g)}{}^{\d}~\nn\\
-\frac{i}{2}\s^{g}{}_{(\a\b|}\s^{mn}{}_{|\g)}{}^{\d}[L^{(2)}{}_{gmn}+\frac{1}{4}A^{(2)}{}_{gmn}]=0
~~\eea
 From which we read

\bea T^{(2)}{}_{\g g}{}^{\d} =-\frac{\g}{2}[\nabla_{\g
}A^{(1)}{}_{gef}]T^{(0)}{}_{ef}{}^{\d} + 2R^{(1)}{}_{\g
gef}T^{(0)}{}_{ef}{}^{\d}\eea

And

\bea R^{(2)}{}_{ \a\b de}
 =-~\frac{i\g}{12}\s^{pqref}{}_{\a\b}A^{(1)}{}_{pqr}R_{ef de}-
 2i\s^{g}{}_{\a\b}{}[L^{(2)}{}_{gde}+\frac{1}{4}A^{(2)}{}_{gde}]
~\eea

\section{Curvature for $R^{(2)}{}_{\la gde}$}

 We need to solve the curvature that
will give $R^{(2)}{}_{\la gde}$. We have the curvature Bianchi
identity which we need to solve at second order as follows

\bea
T_{(\a\b|}{}^{\la}R_{|\g)\la}{}_{de}~-~T_{(\a\b|}{}^{g}R_{|\g)}{}_{gde}~
-~\nabla_{(\a|}R_{\b\g)}{}_{de}~=~0\eea

Here we must consider second order contributions from the spinor
derivative $\nabla_{(\a|}R_{|\b\g)}{}_{de}$. We write the full
curvature to second order for clarity

\bea
R_{\b\g}{}_{de}=~-2i\s^{g}{}_{\a\b}\Pi'{}_{gde}+~\frac{i}{24}\s^{pqr}{}_{de}{}_{\a\b}A^{(1)}{}_{pqr}
-~\frac{i\g}{12}\s^{pqrab}{}_{\a\b}A^{(1)}{}_{pqr}R_{deab} \eea

Where $\Pi'$ is the modified $\Pi$, but is any case is of the
solvable form.

\bea
\Pi'=[L^{(0)}{}_{gde}+L^{(1)}{}_{gde}+L^{(2)}{}_{gde}-\frac{1}{4}A^{(1)}{}_{gde}+\frac{1}{4}A^{(2)}{}_{gde}]\eea

With hindsight and in order to eliminate an apparently intractable
term we begin by making the following observations. In equation (65)
we found $T^{(2)}{}_{\a\b}{}^{\la}$, hence we also encountered the
quantity

 \bea
\s_{d(\a|\la}T^{(2)}{}_{\a\b}{}^{\la}~=~-\frac{i\g}{12}\s_{d(\a|\la}\s^{pqref}{}_{|\b\g)}A^{(1)}{}_{pqr}T^{(0)}_{ef}{}^{\la}
\eea Using the torsion, equation (20), we can write

\bea \s_{d(\a|\la}T^{(2)}{}_{|\b\g)}{}^{\la}~
=-~~\s^{g}{}_{(\a\b|}T^{(2)}{}_{d|\g) g}~~\nn\\
+~iT^{(0)}{}_{(\a\b|}{}^{\la}T^{(2)}{}_{|\g)\la}{}^{d}~
-i\nabla_{(\a|}T^{(2)}{}_{|\b\g)}{}^{d} \eea

\noindent Using the second order torsion results which we found,
(65) and (105),  then gives

 \bea +\s_{d(\a|\la}T^{(2)}_{|\b\g)}{}^{\la}~=
+~\s^{g}{}_{(\a\b|}T^{(2)}{}_{d|\g) g}~~\nn\\
~~+\frac{\g}{6}T^{(0)}{}_{(\a\b|}{}^{\la}[\s^{pqref}{}_{|\g)\la}
H^{(0)}{}_{def}A^{(1)}{}_{pqr}]~ -~\frac{\g}{6}\s^{pqref}{}_{(\a\b|}
H^{(0)}{}_{def}[\nabla_{|\g)}A^{(1)}{}_{pqr}]\nn\\
-~\frac{\g}{6}\s^{pqref}{}_{(\a\b|}
[\nabla_{|\g)}H^{(0)}{}_{def}]A^{(1)}{}_{pqr} \nn\\
\eea

Now applying our key equation, (39) to (112) gives

 \bea +\s_{d(\a|\la}T^{(2)}_{|\b\g)}{}^{\la}~=
+~\s^{g}{}_{(\a\b|}T^{(2)}{}_{d|\g) g}~~\nn\\
~+~\g\s^{g}{}_{(\a\b|}[\nabla_{|\g)}A^{(1)}{}_{gef}]H^{(0)}{}_{d}{}^{ef}
-4\g\s^{g}{}_{(\a\b|}R^{(1)}{}_{|\g)gef}H^{(0)}{}_{d}{}^{ef}\nn\\~~\eea

We now substitute in our result for
$\s^{g}{}_{(\a\b|}T^{(2)}{}_{d|\g) g}$, (78), into (113) to obtain
cancelations and the simple result

\bea \s_{d(\a|\la}T^{(2)}{}_{|\b\g)}{}^{\la}~=
-~\frac{i\g}{6}\s^{g}{}_{(\a\b|}\s^{pqre}{}_{g}{}_{|\g)\f}A^{(1)}{}_{pqr}T^{(0)}{}_{de}{}^{\f}\nn\\
=-\frac{i\g}{12}\s_{d(\a|\la}\s^{pqref}{}_{|\b\g)}A^{(1)}{}_{pqr}T^{(0)}_{ef}{}^{\la}
\nn\\~~\eea

From which we extract the following result. \bea
\s^{pqref}{}_{(a\b|}\s_{d|\g)|\f}A^{(1)}{}_{pqr}M_{ef}{}^{\f}
=2\s^{g}{}_{(\a\b|}\s^{pqre}{}_{g}{}_{|\g)\f}A^{(1)}{}_{pqr}M_{de}{}^{\f}\nn\\
\eea From this we deduce a hitherto unknown identity, albeit
indirectly, where $M_{de}{}^{\f}$ is anti-symmetric in 'd' and 'e'.
Using our second order torsion and curvature results, (65), (69),
and (106), the full curvature at second order becomes

 \bea ~T^{(0)}{}_{(\a\b|}{}^{\la}[ -~\frac{i\g}{12}\s^{pqrab}{}_{|\g)\la}A^{(1)}{}_{pqr}R^{(0)}{}_{abde}]~
 +~[-~\frac{i\g}{12}\s^{pqrab}{}_{(\a\b|}A^{(1)}{}_{pqr}T_{ab}{}^{\la}R^{(0)}{}_{|\g)\la}{}_{de}\nn\\
~-~i\s^{g}{}_{(\a\b|}R^{(2)}{}_{|\g)}{}_{gde}~+~\frac{i\g}{6}\s^{pqrab}{}_{(\a\b|}
H^{(0)}{}^{g}{}_{ab}A^{(1)}{}_{pqr}]R^{(0)}{}_{|\g)}{}_{gde}\nn\\
+~\frac{i\g}{12}\s^{pqrab}{}_{(\a\b|}[\nabla_{|\g)|}A^{(1)}{}_{pqr}]R^{(0)}{}_{abde}
+~\frac{i\g}{12}\s^{pqrab}{}_{(\a\b|}A^{(1)}{}_{pqr}[\nabla_{|\g)}R^{(0)}{}_{abde}]~\nn\\
-~\frac{i}{24}\s^{pqr}{}_{de}{}_{(\a\b|}[\nabla_{|\g)}A^{(1)}{}_{pqr}]
+~2i\g\s^{g}{}_{(\a\b|}[\nabla_{|\g)}\Pi'{}_{gde}] =~0 \eea

Using (39) again gives two more solvable terms

\bea
~+\frac{i\g}{12}T^{(0)}{}_{(\a\b|}{}^{\la}\s^{pqr}{}_{ab|\g)\la}A^{(1)}{}_{pqr}R^{(0)}{}^{ab}{}_{de}-
\frac{i\g}{12}\s^{pqrab}{}_{(\a\b|}R^{(0)}{}_{abde}\nabla_{|\g)}A^{(1)}{}_{pqr}\nn\\
=+\frac{i\g}{2}\s^{g}{}_{(\a\b|}[\nabla_{|\g)}A^{(1)}{}_{gab}]R^{(0)}{}^{ab}{}_{de}-
2i\g\s^{g}{}_{(\a\b|}R^{(1)}{}_{|\g)
g}{}^{ab}R^{(0)}{}_{abde}\nn\\\eea

This reduces  (116) to
 \bea~~+\frac{i\g}{2}\s^{g}{}_{(\a\b|}\nabla_{|\g)}A^{(1)}{}_{gab}R^{(0)}{}^{ab}{}_{de}-
2i\g\s^{g}{}_{(\a\b|}R^{(1)}{}_{|\g) gab}R^{(0)}{}_{abde}\nn\\
-~\frac{i\g}{12}\s^{pqrab}{}_{(\a\b|}A^{(1)}{}_{pqr}T_{ab}{}^{\la}R^{(0)}{}_{|\g)\la}{}_{de}\nn\\
~-~i\s^{g}{}_{(\a\b|}R^{(2)}{}_{|\g)}{}_{gde}~+~\frac{i\g}{6}\s^{pqrab}{}_{(\a\b|}
H^{(0)}{}^{g}{}_{ab}A^{(1)}{}_{pqr}R^{(0)}{}_{|\g)}{}_{gde}\nn\\
+~\frac{i\g}{12}\s^{pqrab}{}_{(\a\b|}A^{(1)}{}_{pqr}[\nabla_{|\g)}R^{(0)}{}_{abde}]~\nn\\
-~\frac{i}{24}\s^{pqr}{}_{de}{}_{(\a\b|}[\nabla_{|\g)}A^{(1)}{}_{pqr}]
+~2i\g\s^{g}{}_{(\a\b|}[\nabla_{|\g)}\Pi'{}_{gde}]=~0\nn\\
\eea

We now list the sigma five terms separately.

\bea +\frac{i\g}{12}\s^{pqrab}{}_{(\a\b|}A^{(1)}{}_{pqr}[
-T^{(0)}{}_{ab}{}^{\la}R^{(0)}{}_{\la|\g)}{}_{de}~
+~2H^{(0)}{}_{ab}{}^{g}R^{(0)}{}_{|\g)}{}_{gde}~+~\nabla_{|\g)}R^{(0)}{}_{abde}]
\eea

\bea =+\frac{i\g}{12}\s^{pqrab}{}_{(\a\b|}A^{(1)}{}_{pqr}[
-T^{(0)}{}_{ab}{}^{\la}R^{(0)}{}_{\la|\g)}{}_{de}~-~T^{(0)}{}_{ab}{}^{g}R^{(0)}{}_{\g}{}_{gde}~
+~\nabla_{|\g)}R^{(0)}{}_{abde}] \eea

 We have the Bianchi Identity

\bea \nabla_{\a}R_{abde}~-~T_{\a [a|}{}^{X}R_{X}{}_{|b]d e}~
-~T_{ab}{}^{X}R_{X}{}_{\a de}~+~\nabla_{[a|}R_{|b]\a d e}=0 \eea

The second term on the LHS of  (121) is zero at zeroth order. Hence
we have as follows,

\bea
+\frac{i\g}{12}\s^{pqrab}{}_{(\a\b|}A^{(1)}{}_{pqr}[\nabla_{|\g)}R^{(0)}{}_{abde}]=\nn\\
+\frac{i\g}{12}\s^{pqrab}{}_{(\a\b|}A^{(1)}{}_{pqr}[+T^{(0)}{}_{ab}{}^{\la}R^{(0)}{}_{\la|\g)}{}_{de}~
-~T^{(0)}{}_{ab}{}^{g}R^{(0)}{}_{|\g)}{}_{gde}~-~2\nabla_{a}R^{(0)}{}_{b|\g)de}]
\eea

 Substituting (122) into (118) gives

\bea~+\frac{i\g}{2}\s^{g}{}_{(\a\b|}[\nabla_{|\g)}A^{(1)}{}_{gab}]R^{(0)}{}^{ab}{}_{de}-
2i\g\s^{g}{}_{(\a\b|}R^{(1)}{}_{|\g) gab}R^{(0)}{}_{abde}\nn\\
~-~i\s^{g}{}_{(\a\b|}R^{(2)}{}_{|\g)}{}_{gde}~
+~\frac{i\g}{6}\s^{pqrab}{}_{(\a\b|}A^{(1)}{}_{pqr}[\nabla_{a}R^{(0)}{}_{|\g)bde}+
2H^{(0)}{}_{ab}{}^{g}R^{(0)}{}_{|\g)gde}]
\nn\\-~\frac{i}{24}\s^{pqr}{}_{de}{}_{(\a\b|}[\nabla_{|\g)}A^{(1)}{}_{pqr}]
+~2i\g\s^{g}{}_{(\a\b|}[\nabla_{|\g)}\Pi'{}_{gde}]=0\eea

Now consider the sigma five terms in (123). Using our new result
result (115) allows for solving these terms, and we obtain \bea
+~\frac{\g}{6}\s^{pqrab}{}_{(\a\b|}A^{(1)}{}_{pqr}[\nabla_{a}R^{(0)}{}_{|\g)bde}+
2H^{(0)}{}_{ab}{}^{g}R^{(0)}_{|\g)gde}]\nn\\
=~\frac{\g}{6}\s^{pqrab}{}_{(\a\b|}A^{(1)}{}_{pqr}[\s_{[d|g)\f}\{\nabla_{a}T^{(0)}{}_{b|e]}{}^{f}
+2H^{(0)}{}_{ab}{}^{c}T^{(0)}{}_{c|e]}{}^{\f}\}] \nn\\
= ~\frac{\g}{3}\s^{g}{}_{(\a\b|}\s^{pqra}{}_{g}{}_{|\g)\f}
A^{(1)}{}_{pqr}[\{\nabla_{[d|}T^{(0)}{}_{a|e]}{}^{f}
+2H^{(0)}{}_{[d|a}{}^{c}T^{(0)}{}_{c|e]}{}^{\f}\}] \eea

Hence we obtain

\bea~ ~i\s^{g}{}_{(\a\b|}R^{(2)}{}_{|\g)}{}_{gde}~=
+\frac{i\g}{2}\s^{g}{}_{(\a\b|}[\nabla_{|\g)}A^{(1)}{}_{gab}]R^{(0)}{}^{ab}{}_{de}-
2i\g\s^{g}{}_{(\a\b|}R^{(1)}{}_{|\g) gab}R^{(0)}{}_{abde}\nn\\
+~\frac{\g}{3}\s^{g}{}_{(\a\b|}\s^{pqra}{}_{g}{}_{|\g)\f}
A^{(1)}{}_{pqr}[\{\nabla_{[d|}T^{(0)}{}_{a|e]}{}^{\f}
+2H^{(0)}_{[d|a}{}^{c}T^{(0)}{}_{|c|e]}{}^{\f}\}]
\nn\\-~\frac{i}{24}\s^{pqr}{}_{de}{}_{(\a\b|}[\nabla_{|\g)}A^{(1)}{}_{pqr}]
+~2i\g\s^{g}{}_{(\a\b|}[\nabla_{|\g)}\Pi'{}_{gde}]=0\nn\\\eea

We now look at the remaining unsolved term
$-~\frac{i}{24}\s^{pqr}{}_{de}{}_{(\a\b|}[\nabla_{|\g)}A^{(1)}{}_{pqr}]$.

This term looks as though it will pose a serious problem. This term
cannot be manipulated into a solvable term because of the placement
of the free indices. Using the results found in \cite{4} we have

 \bea-~\frac{i}{24}\s^{pqr}{}_{de}{}_{(\a\b|}[\nabla_{|\g)}A^{(1)}{}_{pqr}]=\nn\\
+\frac{\g}{(12)(24)}\s^{pqr}{}_{de(\b\g|}\s_{pqr
\eps\tau}T^{(0)}{}^{\eps}{}_{kl}\s^{mns}{}^{\f}{}^{\tau}[H^{(0)}{}^{kl}{}_{g}\s^{g}{}_{|\a)}{}_{\f}A^{(1)}{}_{mns}\nn\\
-\s_{k}{}_{|\a)}{}_{\f}(\nabla_{l}A^{(1)}{}_{mns})] \eea

Using the sigma matrix identities as given in ref. [3], it can be
shown that these two terms cannot be written in the solvable form,
that is with the same structure as
$\s^{g}{}_{(\a\b|}R^{(2)}{}_{|\g)}{}_{gde}$. Hence we look at the
origin of these terms. For the derivative of $T_{kl}{}^{\tau}$ we
have the following Bianchi identity.

\bea
\nabla_{\g}T_{kl}{}^{\tau}=T_{\g[k|}{}^{\la}T_{\la|l]}{}^{\tau}+
T_{\g[k}{}^{g}T_{g|l]}{}^{\tau}+T_{kl}{}^{\la}T_{\la
\g}{}^{\tau}+T_{kl}{}^{g}T_{g\g}{}^{\tau}-\nabla_{[k|}T_{|l]\g}{}^{\tau}
-R_{kl\g}{}^{\tau}\eea

At first order this reduces to

\bea \nabla_{\g}T_{kl}{}^{\tau}{}^{Order (1)}=
T^{(1)}{}_{\g[k|}{}^{\la}T^{(0)}{}_{\la|l]}{}^{\tau}-\nabla_{[k|}T_{|l]\g}{}^{\tau}{}^{Order(1)}
-R^{(1)}{}_{kl\g}{}^{\tau}\eea

In references \cite{2} and \cite{4} it appears that $R^{(1)}{}_{kl
\g}{}^{\tau}$ was set to zero. With the form of the curvature
$R_{\a\b de}$ and this choice of super current supertensor $A_{abc}$
we will always be led to the term
$\frac{i}{24}\s^{pqr}{}_{de}{}_{(\a\b|}\nabla_{|\g)}A^{(1)}{}_{pqr}{}^{(order2)}$
because of the the spinor derivative in the Bianchi identity (107)
as given in (112). This term is not reducible as we require so it
must be incorporated into this curvature. Hence we identify the
following curvature at first order.

\bea R^{(1)}{}_{kl\g}{}^{\tau}=\frac{1}{48}
[2H^{(0)}{}_{kl}{}_{g}\s^{g}{}_{\g
\la}\s^{pqr\la\tau}A^{(1)}{}_{pqr}
-\s_{[k|\g\la}\s^{pqr\la\tau}{}(\nabla_{|l]}A^{(1)}{}_{pqr})]\eea

This result was not arrived at in \cite{2} and \cite{4} and will
have consequences for the application of even the first order
results.

The second order form of this curvature is already solved in the
Bianchi identity (127). All the quantities in this Bianchi identity
are known. Hence it can be written in full in a later review. It is
the role of this paper simply to arrive at the second order
solution, and to overcome obstacles to obtaining this solution.
\section{The Super-Current}
At first the author believed that finding the supercurrent tensor
$A^{(2)}{}_{abc}$ would result in closing the curvature identity
(107). However that was before equation (115) was constructed and
the before the significance of the (128) was realized. Thence
finding the supercurrent $A^{(2)}{}_{abc}$ is nothing more than
applying a condition available from the conventional constraints.
The starting point in references \cite{2} and \cite{4} were the
conventional constraints as listed in \cite{2}. Among these
constraints we have

\bea T_{\a b}{}^{\d}=\frac{1}{48}\s_{b\a \la}\s^{pqr
\la\d}A_{pqr}\eea The choice of

\bea A_{pqr}=-i\g \s_{pqr \eps \tau}T_{kp}{}^{\eps}T^{kp \tau}\eea
 was made for on
shell conditions, [2]. However this is a conventional constraint and
therefore it can be imposed to all orders. Hence we can use this
result. We have found $T_{\a b}{}^{\d}$ and it is given in in
equation (105). Hence we can solve for $A^{(2)}{}_{pqr}$. No
modification to this super-current was required to close the
identities other than this. Hence we use a suitable inverting
operator along with our results (105) and (80) to obtain after som
calculation

 \bea A^{(2)}{}_{gef}=-\frac{1}{20}\s_{gef \g\la}\s^{b
\la \f}T^{(2)}{}_{\f b}{}^{\g} \eea.

\bea =\frac{\g}{20}\s_{gef \g\la}\s^{b \la
\f}[\nabla_{\f}(\frac{1}{4}A^{(1)}{}_{bmn}
-2L^{(1)}{}_{bmn})]T^{(0)mn \g}\eea

\section{Conclusions}

We have solved the non-minimal case of string corrected
supergravity, for D=10, N=1. This theory is believed to be the low
energy realization of string theory. We found a procedure for
solving the Bianchi identities to this order and thus maintained
manifest supersymmetry to that order. Our solution required the
intricate derivation of equation (39), which we used in conjunction
with several other results and observations. In particular we had to
form an Ansatz for the so called X tensor which would be consistent
within all sectors of the Bianchi identities. The Ansatz that we
found achieved this result. Hence we found a mechanism which allows
for closure of the H sector Bianchi identities and solved a problem
that had existed for many years. We also solved the Bianchi
identities in the torsion and curvature sectors at each dimension,
showing consistency of our set of results. To achieve this in full,
progress was held up as it was necessary to derive yet another
identity, (115), which facilitated the elimination of otherwise
unsolvable terms. Furthermore we also had to observe that, in
contradiction to the results reported in ref. [2] and [4], we had to
show that the curvature $R^{(1)}{}_{ab \g}{}^{\d}$ was not in fact
zero. These observations were not at all immediately transparent. We
found $R^{(1)}{}_{ab \g}{}^{\d}$ to be given by equation (133). We
noted that the second order contribution $R^{(2)}{}_{ab \g}{}^{\d}$,
can be found already form the Bianchi identity (127) by direct
substitution of our already found results at second order. It would
simply be a long expression.

 We saw how the X tensor was necessary for achieving consistent closure
 of the Bianchi identities and our candidate for this tensor, which
 succeeds in doing this, did not require the contribution of a second possible part,
 $Y_{pqrdef}$. Such a contribution would appear in fact to
result in failure to close in the H sector. Hence we feel that out
choice is in fact the unique result.

With $H^{(2)}{}_{\a\b \d}$ set to zero we obtained \bea H^{(2)}{}_{d
\a\b}~=~\s_{\a\b}{}^{g}[{8i\g}H^{(0)}{}_{def}L^{(1)}{}_{g}{}^{ef}~
-~i\g H^{(0)}{}_{def}A^{(1)}{}_{g}{}^{ef}]~-~
\frac{i\g}{12}\s^{pqref}{}_{\a\b}[H^{(0)}{}_{def}A^{(1)}{}_{pqr}]\nn\\
\eea

$H^{(2)}{}_{\a b g}$ must be extracted from (66). We use the
following operator, $\hat{O}$,  to obtain the symmetrized
$H^{(2)}{}_{\a b g}$:

\bea \hat{O}=[\frac{1}{2}\d_{[a}{}^{d}\d_{b]}{}^{g}\d_{\a}{}^{\b}~
-~\frac{1}{12}\eta^{dg}\s_{ab\a}{}^{\b}~+\frac{1}{24}\d_{[a}{}^{(d}\s_{b]}{}^{g)}_{\a}{}^{\b}]
\eea  After a very long calculation we obtain the result

\bea H^{(2)}{}_{\a
ab}=~2\g[\nabla_{\a}(H^{(0)}_{[a|ef}H^{(0)}{}_{|b]}{}^{ef})~
-~\s_{ab\a}{}^{\f}\nabla_{\f}(H^{(0)}{}_{gef}H^{gef})]\nn\\
~+2i\g\s_{[a|\a\f}T_{ef}{}^{\f}\Pi^{(1)}{}_{|b]}{}^{ef}
-~2i\g\s_{ab\a}{}^{\la}\s_{g\la\f}T_{ef}{}^{\f}\Pi^{(1)gef}\nn\\
-~\frac{\g}{6}\s^{g}{}_{[a|\a}{}^{\f}\s_{|b]\la\f}T_{ef}{}^{\la}\Pi^{(1)}{}_{g}{}^{ef}
-~\frac{\g}{6}\s^{g}{}_{[a|\a}{}^{\f}\s_{g\la\f}T_{ef}{}^{\la}\Pi^{(1)}{}_{|b]}{}^{ef}\nn\\
~-~4\g R^{(1)}{}_{\a[a|}{}^{ef}H^{(0)}{}_{|b]ef}+T^{(2)}{}_{\a
ab}\eea

\noindent where $T^{(2)}{}_{\a ab}$ is given in equation (142). In
the case of $H^{(2)}{}_{a b c}$, the Bianchi identity has already
given us the result. From the term $T_{\a\b}{}^{E}H_{Ecd}$ in
equation (5) we isolate an expression of the form \bea
T^{(0)}{}_{\a\b}{}^{g}H^{(2)}{}_{gcd}~=~i\s_{\a\b}{}^{g}H^{(2)}{}_{gcd}~=~M_{\a\b cd}\nn\\
\eea The right hand side contains now known torsions and curvatures.
However they need only be substituted into (137) generating a long
expression. We then use the fact that \bea
\s_{a}{}_{\a\b}\s^{b}{}^{\a\b}~=~-~16\d^{b}{}_{a}\eea and solve for
$H^{(2)}{}_{gcd}$.

We obtained the full set or torsions and and curvatures,

 \bea
T^{(2)}{}_{\a\b}{}^{\la}~=~-~\frac{i\g}{12}\s^{pqref}{}_{\a\b}A^{(1)}{}_{pqr}T_{ef}{}^{\la}
\eea

\bea T^{(2)}{}_{\a\b}{}^{d}~=-~\frac{i\g}{6}\s^{pqref}{}_{\a\b}
H^{(0)}{}^{d}{}_{ef}A^{(1)}{}_{pqr}\eea

\bea T^{(2)}{}_{\g g}{}^{\d} =-\frac{\g}{2}[\nabla_{\g
}A^{(1)}{}_{gef}]T^{(0)}{}_{ef}{}^{\d} + 2R^{(1)}{}_{\g
gef}T^{(0)}{}_{ef}{}^{\d}\eea

Extracting the symmetrized torsion $T^{(2)}{}_{\g gd}$ from (78)
gives

 \bea T^{(2)}{}_{\g ab}=
-\frac{\g}{2}[\nabla_{\g}A^{(1)}{}_{[a|ef}]H^{(0)}{}_{|b]}{}^{ef}+
2\g R^{(1)}{}_{\g [a|ef}H^{(0)}{}_{|b]}{}^{ef}
-\frac{i\g}{12}\s^{pqrg}{}_{[a|\g\la}T^{(0)}{}_{|b]g}{}^{\la}A^{(1)}{}_{pqr}\nn\\
+\s_{ab~\g}{}^{\f}[+\frac{\g}{12}
(\nabla_{\f}A^{(1)}{}_{gef})H^{(0)}{}^{gef}+\frac{\g}{3}R^{(1)}{}_{\f~gef}H^{(0)}{}^{gef}
-\frac{i\g}{72}\s^{pqreg}{}_{\f\la}A^{(1)}{}_{pqr}T^{(0)}{}_{eg}{}^{\la}]\nn\\
+\s_{[a|}{}^{g}{}_{\g}{}^{\f}[-\frac{\g}{2}(\nabla_{\f}A^{(1)}{}_{|b]ef})H^{(0)}{}_{g}{}^{ef}
-\frac{\g}{2}(\nabla_{\f}A^{(1)}{}_{gef})H^{(0)}{}_{|b]}{}^{ef}\nn\\
-\frac{\g}{6}R^{(1)}{}_{\f~|b]ef}H^{(0)}{}_{g}{}^{ef}-\frac{\g}{6}R^{(1)}{}_{\f~gef}H^{(0)}{}_{|b]}{}^{ef}\nn\\
+\frac{i\g}{144}A^{(1)}{}_{pqr}[\s^{pqre}{}_{|b]}{}_{\f\la}T^{(0)}{}_{eg}{}^{\la}
+\s^{pqre}{}_{g~\f\la}T^{(0)}{}_{e|b]}{}^{\la}]]\nn\\
\eea

where \bea
\nabla_{\g}A^{(1)}{}_{gef}=i\g\s_{gef~\eps\tau}T^{(0)}{}_{kp}{}^{\eps}[2T^{(0)}{}^{kp}{}^{
\la}T^{(0)}{}_{\g\la}{}^{\tau}-\frac{1}{2}\s^{mn}{}_{\g}{}^{\tau}R^{(0)}{}^{kp}{}_{mn}]\nn\\
\eea

\bea R^{(2)}{}_{ \a\b de}
 =-~\frac{i\g}{12}\s^{pqref}{}_{\a\b}A^{(1)}{}_{pqr}R_{ef de}-
 2i\s^{g}{}_{\a\b}{}[L^{(2)}{}_{gde}+\frac{1}{4}A^{(2)}{}_{gde}]
~\eea

We also find the adjusted curvature $R{}_{kl\g}{}^{\tau}$. For  $
R^{(2)}{}_{\a\b de}$ we have reduced it to solvable form. After
imposing conditions (133) and (115) on (110) we obtain
 \bea~
~i\s^{g}{}_{(\a\b|}R^{(2)}{}_{|\g)}{}_{gde}~=
\s^{g}{}_{(\a\b|}[\frac{i\g}{2}[\nabla_{|\g)}A^{(1)}{}_{gab}]R^{(0)}{}^{ab}{}_{de}-
2i\g R^{(1)}{}_{|\g)gab}R^{(0)}{}_{abde}\nn\\
+~\frac{\g}{3}\s^{pqra}{}_{g}{}_{|\g)\f}
A^{(1)}{}_{pqr}\{\nabla_{[d|}T^{(0)}{}_{a|e]}{}^{\f}
+2H^{(0)}_{[d|a}{}^{c}T^{(0)}{}_{c|e]}{}^{\f}\}+~2i\g\nabla_{|\g)}\Pi'{}_{gde}]=0\eea

$ R^{(2)}{}_{\a\b de}$ can be extracted from the above result.

Finally we have found the supercurrent $A^{(2)}{}_{abc}$

\bea =\frac{\g}{20}\s_{gef \g\la}\s^{b \la
\f}[\nabla_{\f}(\frac{1}{4}A^{(1)}{}_{bmn}
-2L^{(1)}{}_{bmn})]T^{(0)mn \g}\eea

\section{Appendix I: Conventions}

It has been shown that the Lorentz Chern-Simmons Form, $Q_{ABC}$,
can be defined in superspace for the various dimensionalities as
follows, \cite{1},

\bea Q_{\a\b\g}=\frac{1}{2}\omega_{(\a|}{}^{ef}R_{|\b\g)gf}-
\frac{1}{3}\omega_{(\a|}{}^{ef}\omega_{|\b|}{}^{g}{}_{e}\omega_{|\g)ef}\nn\\
Q_{\a\b c}=\omega_{(\a|}{}^{ef}R_{|\b) cef}+\omega_{c}{}^{ef}R_{\a\b
cef}-\omega_{(\a|}{}^{ef}\omega_{|\b)}{}^{g}{}_{e}\omega_{cgf}\nn\\
Q_{\a bc}=\omega_{(\a|}{}^{ef}R_{b cef}+\omega_{c}{}^{ef}R_{|c]\a
ef}- \omega_{[a|}{}^{ef}\omega_{|c]}{}^{g}{}_{e}\omega_{\a gf}
\nn\\
Q_{abc}=\frac{1}{2}\omega_{[a|}{}^{ef}R_{|bc]ef}-
\frac{1}{3}\omega_{[a|}{}^{ef}\omega_{|b|}{}^{g}{}_{e}\omega_{|c]gf}\eea

D=10, N=1 appears in many formalisms, possibly related by a Weyl
Transformation,[1]. For a review of this theory at zeroth order see
[3]. Hence we require from D=10, N=1 Supergravity the field strength
$G_{ABC}$, (see references \cite{1}, and many references therein).

\bea G_{ABC}=\frac{1}{2}\partial_{[B|}B_{|BC)}\eea or it can be
coupled to a Yang Mills Supermultiplet, \cite{1}

In this paper we work with the modified field strength, $H_{ABC}$,
as defined in equation equation (13) as opposed to $G_{ABC}$ in
reference \cite{2}. We also have the supercurrent supertensor
$A_{ABC}$ as defined in equation (24) for on shell conditions, and
later modified to second order in equation (133). This modification
includes the string tension parameter $\a$. This correction was
first given in \cite{1}.

In this paper we use $\g$ where. \bea \g=const.\a\eea

The field strength $L_{ABC}$ is simply given by

\bea H_{ABC}=-2L_{ABC}\eea.

\noindent We have the curvatures and torsions $T_{AB}{}^{C}$, and
curvatures, $R_{ABCD}$, for their various dimensions
 We also
encounter the super field fields $\f$ and  $\chi$ which is given by

\bea \chi_{\a}=-\frac{1}{2}\nabla_{\a}\f \eea

\subsection{The Sigma Matrix Algebra}

A considerable array of sigma matrix identities can be found in ref.
\cite{3}. For this work we require to know only those listed here.
Details of all conventions are given in \cite{3}. To make our work
self contained however we start form first principles as follows.
Spinors in D=10 space-time dimensions are sixteen component objects.
Let dotted indices be right handed components and un-dotted indices
left handed. Let let space-time indices be in small Roman script and
spinor indices be Greek letters. We deal only with purely left
handed (chiral supergravity) spinors so no dotted indices appear.

The local ten dimensional metric $\eta_{a b}$ has signature $[+,
-,-....]$. The sigma matrices are  therefore 16 by 16 matrices which
satisfy the usual Dirac algebra with anti-commutator as follows

\bea \s_{a \a\b}\s_{b}{}^{\b\g}+\s_{b
\a\b}\s_{a}{}^{\b\g}=-2\eta_{ab}\s_{\a}{}^{\g}\eea

\noindent Hence we have \bea \s_{a \a\b}\s_{b}{}^{\b\g}=-\s_{b
\a\b}\s_{a}{}^{\b\g}-2\eta_{ab}\s_{\a}{}^{\g}\eea Or we have

\bea \s_{a
\a\b}\s_{b}{}^{\b\g}=-\s_{ab\a}{}^{\g}-\eta_{ab}\s_{\a}{}^{\g}\eea

\noindent This defines the object $\s_{ab\a}{}^{\g}$. Similarly

\bea \s_{a \a\b}\s_{bc\g}{}^{\b}=-\s_{abc
\a\g}-\frac{1}{2!}\eta_{a[b|}\s_{|c] \a\g}\eea And so on as we can
build up related products. Hence all the sigma matrices are
anti-symmetric in their vector indices. Sigma matrices with odd
numbers of vector indices are anti-symmetric in their spinor
indices. Symmetrization and anti-symmetrization is defined as
follows.

\bea A_{(a|}A_{|b)}= A_{a}A_{b}+A_{b}A_{a}\eea

\bea A_{[a|}A_{|b]}= A_{a}A_{b}-A_{b}A_{a}\eea

The only identities from \cite{3} which we use in this paper are
listed below. The remaining ones that we use are derived in the
appendix. We have the following identities which we used once or
more

\bea \s_{a \a\b}\s^{b \a\b}=-\d_{a}{}^{b}\eea

\bea \s_{ab
\a}{}^{\b}\s^{cd}{}_{\b}{}^{\a}=-\d_{[a|}{}^{c}\d_{|b]}{}^{d}\eea

\bea \s_{abc
\a\b}\s^{def\a\b}=-\d_{[a|}{}^{d}\d_{b}{}^{e}\d_{|c]}{}^{f}\eea

\bea
\s_{abc\a\b}\s^{abc\g\d}=-8.3!.\d_{[\a|}{}^{\g}\d_{|\b]}{}^{\d}\eea

\bea
\s_{abc\a\b}\s^{abc}{}_{\g\d}=-2.3!.\d^{a}{}_{\a[\g|}\s_{a|\d]\b}\eea

\bea \s_{a \a\b}\s^{a \b\g}=-10\d_{\a}^{\g}\eea

Finally we have the important result \bea \s^{pqr}{}_{ef}{}_{\a\b} =
[\eta^{[p}{}_{e}\eta^{q}_{f}\s^{r]}{}_{\a\b}-\frac{1}{2}\s_{ef}{}_{\a}{}^{\f}\s^{pqr}{}_{\f
\b}] \eea

To complete the set we require also equation (40).

\section{Appendix II}

Using the notation of [6] we define torsions and curvatures as
follows,

 \bea
[\nabla_{A},\nabla_{B}\}=T_{AB}{}^{C}+\frac{1}{2}R_{ABd}{}^{e}M_{e}{}^{d}
\eea

In this paper we required the following,

\bea
T_{(\a\b|}{}^{\la}T_{|\g)\la}{}^{\d}~-~T_{(\a\b|}{}^{g}T_{|\g)g}{}^{\d}
~-~\nabla_{(\a|}T_{|\b\g)}{}^{\d}-\frac{1}{4}R^{(2)}{}_{(\a\b|
de}\s^{de}{}_{|\g)}{}^{\d} ~=~0 \eea

\bea
T_{(\a\b|}{}^{\la}T_{|\gamma)\la}{}^{d}~-~T_{(\a\b|}{}^{g}T_{|\gamma)g}{}^{d}~
-~\nabla_{(\a|}T_{\b\g)}{}^{d}~=~0 \eea

\bea
T_{(\a\b|}{}^{\la}R_{|\g)\la}{}_{de}~-~T_{(\a\b|}{}^{g}R_{|\g)}{}_{gde}~
-~\nabla_{(\a|}R_{\b\g)}{}_{de}~=~0\eea

\bea \nabla_{\a}R_{abde}~=~T_{\a [a|}{}^{E}R_{E}{}_{|b]d e}~
+~T_{ab}{}^{E}R_{E}{}_{\a de}~-~\nabla_{[a|}R_{|b]\a d e} \eea

\section{Appendix III}
A lengthy derivation available from author.

\section{Appendix IV: Calculation of Equation (37)}

A lengthy derivation available from author.

\section{Appendix V: Calculation of Equation (62)}

We have the term

\bea
\frac{+i\g}{24}\s^{pqref}{}_{(\a\b|}A^{(1)}{}_{pqr}[R^{(0)}{}_{|\g)def}~-
~2\nabla_{|\g)}H^{(0)}{}_{def}] ~~ \eea

\bea
~=~\frac{+i\g}{24}\s^{pqref}{}_{(\a\b|}A^{(1)}{}_{pqr}[-i\s_{[e|\g)\f}T^{(0)}{}^{\f}{}_{d|f]}
~-~2\frac{i}{4}\s_{[d|\g)\f}T^{(0)}{}^{\f}{}_{|ef]}] ~~ \eea

\bea
~=~(-i)\frac{+i\g}{24}\s^{pqref}{}_{(\a\b|}A^{(1)}{}_{pqr}[2\s_{e|\g)\f}T^{(0)}{}^{\f}{}_{d
f} ~+~\frac{1}{2}[2\s_{d |\g)\f}T^{(0)}{}^{\f}{}_{ef}~-~4\s_{e
\g)\f}T^{(0)}{}^{\f}{}_{d f}]] ~~ \eea

\bea
~=~\frac{+\g}{24}\s^{pqref}{}_{(\a\b|}A^{(1)}{}_{pqr}\s_{d|\g)\f}T^{(0)}{}^{\f}{}_{\e
f} ~~\eea

Consider

\bea
\s^{pqref}{}_{\a\b}\s_{d\g\f}T^{(0)}{}_{ef}{}^{\f}A^{(1)}{}_{pqr}~~
=[\eta^{[p}_{e}\eta^{q}_{f}\s^{r]}_{\a\b}
~-~\frac{1}{2}\s_{ef(\a|}{}^{\f}\s^{pqr}{}_{|\b)\f}]A^{(1)}{}{}_{pqr}\s_{d\g\la}T^{(0)}{}^{ef}{}^{\la}\nn\\
~\eea

\bea =[6\s^{g}_{\a\b
}A^{(1)}{}_{gef}\s_{d\g\la}T^{(0)}{}^{ef}{}^{\la}
~-~\frac{1}{2}\s_{ef(\a|}{}^{\f}\s^{pqr}{}_{|\b)\f}]A^{(1)}{}{}_{pqr}\s_{d\g\la}T^{(0)}{}^{ef}{}^{\la}\nn\\
~=\la_{1}~+~\la_{2}~~~ \eea

Using the definition of$~~A^{(1)}{}_{gef}~~$ gives

\bea \la_{2}~=~-~\frac{1}{2}\s_{ef(\a|}{}^{\f}\s^{pqr}{}_{|\b)\f}
[-i\g\s{}_{pqr \eps\tau}]\s_{d\g\la}T^{(0)}{}^{ef}{}^{\la}
T^{(0)}{}_{kp}{}^{\eps}T^{(0)}{}^{kp}{}^{\tau} \eea

Using ref. \cite{3} and the appropriate sigma matrix result, gives

\bea =~+~\frac{1}{2}\s_{ef(\a|}{}^{\f} [-12i\s^{g}{}_{|\b)[\eps
|}\s_{g}{}_{|\tau]\f}]\s_{d\g\la}T^{(0)}{}^{ef}{}^{\la}
T^{(0)}{}_{kp}{}^{\eps}T^{(0)}{}^{kp}{}^{\tau} \eea

 With the anti symmetry in $\eps$ and $\tau$ gives a factor of 2,
\bea =~(-12i)\g\s_{ef(\a|}{}^{\f}\s^{g}{}_{|\b)\eps}
\s_{g}{}_{\tau\f}\s_{d\g\la}T^{(0)}{}^{ef}{}^{\la}
T^{(0)}{}_{kp}{}^{\eps}T^{(0)}{}^{kp}{}^{\tau}\nn\\
 \eea
We also have the basic result that
 \bea \s_{ef
\a}{}^{\f}\s_{g}{}_{\tau\f}~ =~-~\eta_{g[e}\s_{f]}{}_{\a
\tau}~-~\s_{gef \a  \tau}\eea Similarly, \bea \s_{ef
\tau}{}^{\f}\s_{g}{}_{\a \f}~ =~-~\eta_{g[e}\s_{f]}{}_{\a
\tau}~-~\s_{gef \tau \a} \eea

Now add the above two equations and use the anti-symmetry in the
spinor indices of  $\s_{gef \tau \a}$ to get

\bea \s_{ef \a}{}^{\f}\s_{g}{}_{\tau\f}~
=~-~2\eta_{g[e}\s_{f]}{}_{\a \tau}~-~\s_{ef \tau}{}^{\f}\s_{g \a \f}
\eea

Hence $\la_{2}$ becomes

 \bea \la_{2}~=~(-12i)\g\s^{g}{}_{(\b|\eps}
[-2\eta_{g[e|}\s_{|f]|\a)\tau}~~-~~\s_{ef\tau}{}^{\f}\s_{g|\a)\f}]\s_{d\g\la}T^{(0)}{}^{ef}{}^{\la}
T^{(0)}{}_{kp}{}^{\eps}T^{(0)}{}^{kp}{}^{\tau} \eea

Noting also the antisymmetry in e and f, we get

\bea =~+48
i\g\eta_{ge}\s_{f(\a|\tau}\s^{g}{}_{|\b)\eps}\s_{d\g\la}T^{(0)}{}^{ef}{}^{\la}
T^{(0)}{}_{kp}{}^{\eps}T^{(0)}{}^{kp}{}^{\tau}\nn\\
~~+~12i\g\s_{ef\tau}{}^{\f}\s_{g}{}_{(\a|\f}\s^{g}{}_{|\b)\eps}\s_{d\g\la}T^{(0)}{}^{ef}{}^{\la}
T^{(0)}{}_{kp}{}^{\eps}T^{(0)}{}^{kp}{}^{\tau} \eea

We have the result,

\bea
\s_{g}{}_{(\a|\f}\s^{g}{}_{|\b)\eps}~=~-~\s_{g}{}_{\a\b}\s^{g}{}_{\eps
\f} \eea

We can also show that can
$[\s^{g}{}_{[\eps|\f}\s_{ef|\tau]}{}_{\f}]~~=~~-2\s_{gef\eps\tau}$

Hence
\bea\la_{2}=~+~48i\g\s_{e(\a|\eps}\s_{f|\b)\tau}\s_{d\g\la}T^{(0)}{}^{ef}{}^{\la}
T^{(0)}{}_{kp}{}^{\eps}T^{(0)}{}^{kp}{}^{\tau} \nn\\
~-~12i\g\s_{g}{}_{\a\b}[\frac{1}{2}\s^{g}{}_{[\eps|\f}\s_{ef|\tau]}{}_{\f}\s_{d\g\la}T^{(0)}{}^{ef}{}^{\la}
T^{(0)}{}_{kp}{}^{\eps}T^{(0)}{}^{kp}{}^{\tau} \eea

\bea
~=~+~48i\g\s_{e(\a|\eps}\s_{f|\b)\tau}\s_{d\g\la}T^{(0)}{}^{ef}{}^{\la}
T^{(0)}{}_{kp}{}^{\eps}T^{(0)}{}^{kp}{}^{\tau}~~\nn\\
+~12i\g\s^{g}_{\a\b}\s_{gef \eps
\tau}\s_{d|\g)\la}T^{(0)}{}^{ef}{}^{\la} \eea

So using this and also the definition of $A^{(1)}{}_{gef}$  we
finally get

\bea
~~=~~+~48~(i)\g\s_{e(\a|\eps}\s_{f|\b)\tau}\s_{d|\g)\la}T^{(0)}{}^{ef}{}^{\la}
T^{(0)}{}_{kp}{}^{\eps}T^{(0)}{}^{kp}{}^{\tau}~~\nn\\
-12\s^{g}_{\a\b}A^{(1)}{}_{gef}\s_{d|\g\la}T^{(0)}{}^{ef}{}^{\la}
\eea

This was the second term in equation (117). Adding to $\la_{1}$
gives \bea
\la_{1}~+~\la_{2}~=~+~48(i)\g\s_{e(\a|\eps}\s_{f|\b)\tau}\s_{d\g\la}T^{(0)}{}^{ef}{}^{\la}
T^{(0)}{}_{kp}{}^{\eps}T^{(0)}{}^{kp}{}^{\tau}\nn\\
~+~(6~-~12)\s^{g}_{\a\b}A^{(1)}{}_{gef}\s_{d\g\la}T^{(0)}{}^{ef}{}^{\la}~~
\eea

Hence introducing the symmetries over $\a, \b, \g$ we get the final
result,
 \bea
\frac{+i\g}{24}\s^{pqref}{}_{(\a\b|}A^{(1)}{}_{pqr}[R^{(0)}{}_{|\g)def}~-
~2\nabla_{|\g)}H^{(0)}{}_{def}] ~~\nn\\
=~~-\frac{\g}{4}\s^{g}_{(\a\b|}A^{(1)}{}_{gef}\s_{d|\g)\la}T^{(0)}{}^{ef}{}^{\la}~~\nn\\
+~4i\g^{2}\s_{e(\a|\eps}\s_{f(\b|\tau}\s_{d|\g)\la}T^{(0)}{}^{ef}{}^{\la}T^{(0)}{}_{kp}{}^{\eps}T^{(0)}{}^{kp}{}^{\tau}
~~\eea

\section{Acknowledgements}
I wish to acknowledge the fact that this work would not exist
without S.J. Gates Jr. who laid the ground work, and to whom I am
also grateful for many comments and criticisms. Also I wish to thank
S. Bellucci for introducing me to the method of Bianchi identities,
for pointing out this problem to me and for checking many of the
calculations.


\begin{thebibliography}{99}


\bibitem{1} S.J. Gates, Jr. and H. Nishino, Nucl. Phys. B291 (1987) 205; ibid.
  Phys. Lett. B173 (1986) 52; ibid. S.J. Gates, Jr. and H. Nishino, Phys. Lett. B157 (1985) 157.

\bibitem{2} S.J. Gates, Jr., A. Kiss, W. Merrell, JHEP 0412 (2004) 047.

\bibitem{3} S.J. Gates, Jr. and S. Vashakidze, Nucl. Phys. B291 (1987) 172.

\bibitem{4} S. Bellucci, D.A. Depireaux and S.J. Gates, Jr., Phys. Lett.
B238 (1990) 315.

\bibitem{5} L. Bonora, M. Bregola, R.D. Auria, P. Fre, K. Lechner, P.
Pasti, I. Pesando, M. Raciti, F. Riva, M. Tonin and D. Zanon, Phys.
Lett. B277 (1992) 306.

\bibitem{6} S. J. Gates, Jr, M.T. Grisaru, M. Rocek and W. Siegal,
"Superspace",(Benjamin/Cummings, Reading MA 1083)

\bibitem{7} B. E. W Nilsson, Phys. Lett. B175 (1986) 319

\bibitem{8} M. B. Green and J. H  Schwarz, Phys. Lett. B149 (1984) 117

\end{thebibliography}
\end{document}